\documentclass[12pt]{article}
\pdfoutput=1
\usepackage{array} 
\usepackage{amssymb,dsfont}
\usepackage{graphics,graphpap}
\usepackage{graphicx}
\usepackage{color}
\usepackage{graphicx}
\usepackage{dcolumn}
\usepackage{epsfig}
\usepackage{epstopdf}
\usepackage{bbm}
\usepackage{amsmath}
\usepackage{amsfonts}
\usepackage{textcomp}
\usepackage{slashed}
\usepackage{subcaption}
\usepackage{setspace}
\usepackage{slashed}
\usepackage{multirow}
\usepackage{footnote}
\makesavenoteenv{table}
\usepackage{url}
\usepackage{slashed}
\usepackage{mathtools}
\newcommand*\diff{\mathop{}\!\mathrm{d}}

\usepackage{tikz}
\definecolor{dunkelgruen}{rgb}{0,0.7,0}
\definecolor{dunkelblau}{rgb}{0,0,0.7}

\usepackage{xcolor}
\definecolor{red}{rgb}{1,0,0}
\definecolor{purple}{rgb}{0.5,0,0.5}
\definecolor{blue}{rgb}{0,0,1}

\usepackage{simplewick}

\newcommand{\beq}{\begin{eqnarray}}
\newcommand{\eeq}{\end{eqnarray}}

\newcommand{\bmp}{\noindent\begin{minipage}{16cm}}
\newcommand{\emp}{\end{minipage}\vskip 7mm} 

\def\drawbox#1#2{\hrule height#2pt
        \hbox{\vrule width#2pt height#1pt \kern#1pt
              \vrule width#2pt}
              \hrule height#2pt}

\def\Asym#1#2{\vcenter{\vbox{\drawbox{#1}{#2}
              \kern-#2pt 
              \drawbox{#1}{#2}}}}

\def\simge{\mathrel{%
   \rlap{\raise 0.511ex \hbox{$>$}}{\lower 0.511ex \hbox{$\sim$}}}}

\def\simle{\mathrel{
   \rlap{\raise 0.511ex \hbox{$<$}}{\lower 0.511ex \hbox{$\sim$}}}}

\def\s#1{\setbox0=\hbox{$#1$}%
\rlap{\ifdim\wd0>.7em\kern.22\wd0\else\kern.1\wd0\fi /}#1}

\definecolor{purple}{rgb}{0.5,0,0.5}

\usepackage{cite}

\begin{document}

\begin{titlepage}
\title{\vspace*{-2.0cm}
\hfill {\small MPP-2016-334}\\[20mm]
\vspace*{-1.5cm}
\bf\Large
$\boldsymbol{\mu^-}$-- $\boldsymbol{e^+}$ Conversion from Short-Range Operators \\[5mm]\ \vspace{-1cm}}

\author{
Tanja Geib\thanks{email: \tt tgeib@mpp.mpg.de}~~~and~~Alexander Merle\thanks{email: \tt amerle@mpp.mpg.de}
\\ \\
{\normalsize \it Max-Planck-Institut f\"ur Physik (Werner-Heisenberg-Institut),}\\
{\normalsize \it F\"ohringer Ring 6, 80805 M\"unchen, Germany}
}
\date{\today}
\maketitle
\thispagestyle{empty}

\begin{abstract}
\noindent
We present a detailed discussion of the lepton flavour and number violating conversion of bound muons into positrons. This process is a viable alternative to neutrinoless double beta decay and, given that experiments on ordinary $\mu^-$-- $e^-$ conversion are expected to improve their sensitivities by several orders of magnitude in the coming years, we can also assume the limit on $\mu^-$-- $e^+$ conversion to improve by roughly the same factor. We discuss how new physics at a high scale can lead to short-range contributions to this conversion process and we present one explicit case in great detail (the single one for which the corresponding nuclear matrix element is presently known). The main goal of our discussion is to make the respective computation accessible to the particle physics community, so that promising models can be investigated while the nuclear physics community can simultaneously advance the computation of nuclear matrix elements. Given the progress to be expected on the experimental side, it may even be possible that lepton number violation in the $e\mu$-sector is discovered by $\mu^-$-- $e^+$ conversion before neutrinoless double beta decay can show its existence in the $ee$-sector.
\end{abstract}

\end{titlepage}

\section{\label{sec:intro}Introduction}

The Standard Model (SM) of particle physics is an almost perfect description of the smallest building blocks we know of the Universe. With the only exception of neutrino oscillations~\cite{Olive:2016xmw} (and possibly the anomalous magnetic moment of the muon~\cite{Lindner:2016bgg}), the SM passes all experimental tests. We can turn the logic round, too, and instead derive predictions from the SM which we can test. Among these predictions are the absence of lepton flavour and number violation (abbreviated LFV and LNV, respectively), arising from an accidental symmetry. Experimentally, while LFV is in fact already proven by neutrino oscillations, LNV seems to be more elusive.

Nevertheless, LNV is something particle theorists strongly expect to exist. While the SM Lagrangian seems to conserve lepton number, it does in fact only do so at the perturbative level: one can show that -- even within the SM -- non-perturbative processes exist which violate lepton number~\cite{'tHooft:1976up,Klinkhamer:1984di}. Thus, this quantum number is not sacrosanct. The notion of LNV being something to naturally occur is supported by the effective operator of lowest non-renormalisable dimension, the Weinberg operator~\cite{Weinberg:1979sa}, violating lepton number, too. Thus, as to be expected, any New Physics scenario realising the Weinberg operator does indeed exhibit LNV.

Still it is hard to look for any sign of LNV in an experiment, due to the corresponding processes only having very small rates. The most promising and most intensely investigated process is probably neutrinoless double beta decay ($0\nu\beta\beta$)~\cite{Rodejohann:2010bv}, where a nucleus with atomic number $Z$ and mass number $A$ decays while producing two electrons but not other leptons, $(Z,A) \to (Z+2,A) + 2e^-$, which is clearly an LNV transition. Experiments such as EXO-200~\cite{Albert:2014awa}, KamLAND-Zen~\cite{Gando:2012zm}, or GERDA~\cite{Agostini:2016iid} have been able to push the limits on the lifetimes of isotopes potentially undergoing $0\nu\beta\beta$ to values above $10^{25}$~yrs.

Limits on other LNV processes like kaon decays (e.g.\ NA48~\cite{Massri:2016dff}: BR$(K^\pm \to  \pi^\mp \mu^\pm \mu^\pm) <8.6\cdot 10^{-11}$ at 90\%~C.L.), $B$-meson decays (BaBar~\cite{Lees:2011hb}: BR$(D^+ \to K^- e^+ \mu^+) <1.9\cdot 10^{-6}$ at 90\%~C.L.; BELLE~\cite{Seon:2011ni}: BR$(B^+ \to D^- e^+ \mu^+) <1.8\cdot 10^{-6}$ at 90\%~C.L.), or $\tau$ decays (BELLE~\cite{Miyazaki:2012mx}: BR$(\tau^- \to e^+ \pi^- \pi^-) <2.0\cdot 10^{-8}$ at 90\%~C.L.), cannot compete with $0\nu\beta\beta$. However, note that some processes do not only exhibit LNV but also LFV. Thus, they are worth investigating, since even for the simple case of light Majorana neutrinos, the $m_{ee}$ element of the mass matrix can be subdominant compared to $m_{e\mu/e\tau}$~\cite{Merle:2006du}.

In a recent letter~\cite{Geib:2016atx}, we have pointed out that the process $\mu^-$-- $e^+$ conversion (note the positron in the final state!) could be an interesting route to pursue, as this process would also exhibit both LFV and LNV at the same time. This process had already been theoretically proposed~\cite{Pontecorvo:1967fh,Kisslinger:1971vw,Shuster:1973se} and experimentally studied~\cite{Bryman:1972rf,Abela:1980rs,Badertscher:1981ay,Burnham:1987gr,Ahmad:1988ur,Kaulard:1998rb,Bertl:2006up} in the past, however, new advances on the related LFV-only process of $\mu^-$-- $e^-$ conversion may push the limits on \emph{both} processes by about five orders of magnitude within the coming years~\cite{Raidal:2008jk,Barlow:2011zza}.

As we pointed out in~\cite{Geib:2016atx}, advances are necessary on different frontiers: particle, nuclear, and experimental physics. Indeed, at least the particle physics community seems to have picked up our motivational letter well, with studies of interpretations of a detection in what concerns the flavour space~\cite{Heeck:2016xwg}, of several types of effective operators~\cite{Berryman:2016slh}, and of the complementarity to LFV processes with muons~\cite{Crivellin:2016ebg} appearing shortly after our work.

However, what has been unavailable is a detailed computation of $\mu^-$-- $e^+$ conversion on a level accessible to particle physicists. In this work, we try to make the first step to remedy the situation by presenting a detailed computation of the process when based on the effective operator with the coefficient $\epsilon_3^{xyz}$, where $x, y, z \in \{ L, R \}$ denote chiralities, cf.\ Eq.~\eqref{eq:short-range} and Ref.~\cite{Geib:2016atx}. For illustrative purposes, we will constantly relate this operator to concrete realisations of New Physics scenarios, like mediation by doubly charged scalars~\cite{Chen:2006vn,King:2014uha,Geib:2015tvt,Geib:2015unm}, $R$-parity violating supersymmetry~\cite{Barbier:2004ez,Faessler:1997db}, or heavy right-handed neutrinos~\cite{Domin:2004tk}. Note that some results of~\cite{Domin:2004tk} for the case of heavy right-handed neutrinos carry over to our more general computation, which is particularly true for the nuclear matrix elements (NMEs). However, we add several decisive bits needed for the important results of Ref.~\cite{Domin:2004tk} to be used by particle physicists: 1.)~we provide a guideline on how to realise effective operators by concrete New Physics scenarios; 2.)~we provide the tools to compare different particle physics models to each other, which is the key to understanding which settings could be constrained by $\mu^-$-- $e^+$ conversion~\cite{Geib:2016atx}; 3.)~last but not least, we provide a much more explicit computation than presented in Ref.~\cite{Domin:2004tk}, which will make the technical aspects easier to grasp. Thus, at least for the one effective operator for which NMEs have already been computed, we will make it understandable which elements go into the computation. Should more NME computations arise from the nuclear physics side and should more effective operators be investigated from the particle physics community, with an eye on the comparison between different New Physics scenarios, the present paper will provide the glue necessary to connect these efforts.

This paper is structured as follows. In Sec.~\ref{sec:short-range} we introduce the effective operator language for $\mu^-$-- $e^+$ conversion, which forms the basis for our discussion. The main computation is laid out in Sec.~\ref{sec:rate-explicit}, where we derive the decay rate for $\mu^-$-- $e^+$ induced by $\epsilon_3^{xyz}$ in sufficient detail to enable the reader to reproduce our results. In Sec.~\ref{sec:matching} we show how to map particle physics models to the operator $\epsilon_3^{xyz}$, which is the key to understanding how experimental bounds constrain the possibilities for physics beyond the SM. We conclude in Sec.~\ref{sec:conc}. To make the text as accessible as possible, we have postponed technical aspects to the appendices. Therefore, App.~\ref{app:A} is dedicated to explaining the differences in our notation compared to that of Ref.~\cite{Domin:2004tk}, App.~\ref{app:B} is devoted to detail on how to handle the many spins appearing in the computation, and App.~\ref{app:C} lists all Feynman rules used.

\section{\label{sec:short-range}Possible short-range operators}

We start by discussing the possible short-range contributions to the LNV and LFV $\mu^-$-- $e^+$ conversion. While this discussion had already been touched in Ref.~\cite{Geib:2016atx}, we will here focus a bit more on the technical aspects, in particular when performing the matching of concrete models to the effective operator coefficients.

In order to consider the short-range contributions to the $\mu^-$-- $e^+$ conversion within a general framework, we turn to an effective field theory treatment. Hence, the bound muon and the positron interact with the quarks inside the nucleus via point-like vertices. Due to the charge flow, we can thus imagine the process as having one muon $\mu^-$ and two up-quarks $u$ as ingoing particles and one positron $e^+$ and two down-quarks $d$ as outgoing, all of which being connected via a ``big'' effective vertex. 

We restrict ourselves to the lowest dimensional short-range operators which have dimension~9.\footnote{We will not consider long-range operators in the following, i.e., we consider models in which the New Physics contribution only arises at high energies. However, in principle, the long-range contributions could be parametrised in a similar manner, see~\cite{Pas:1999fc} for a thorough discussion for the case of $0\nu\beta\beta$.} Thus, our effective Lagrangian will consist of combinations of two hadronic currents $J$ and one leptonic current $j$, with a prefactor $G_F^2/m_p$ of mass dimension $(-5)$ to balance out the mass dimensions. Note that the factor $G_F^2$ is motivated by the $W$-bosons that are often present in such a transition. The strength of these vertices will be parametrised by dimensionless coefficients $\epsilon^{xyz}_a$, which are labeled by the index $a$ and whose superscript $xyz$ indicates the currents' chiralities involved in the operators.

Let us now write down the most general short-range Lagrangian, which can be done analogously to $0\nu\beta\beta$ ~\cite{Pas:2000vn}. Taking into account Lorentz invariance, it is given by:
\begin{eqnarray}
 &&\mathcal{L}_\text{short-range}^{\mu e} = \frac{G_F^2}{2 m_p} \sum_{x,y,z = L,R}\big[ \epsilon_1^{xyz} J_x J_y j_z + \epsilon_2^{xyz} J_x^{\nu \rho} J_{y,\nu \rho} j_z + \epsilon_3^{xyz} J_x^\nu J_{y,\nu} j_z + \epsilon_4^{xyz} J_x^\nu J_{y,\nu \rho} j_z^\rho\nonumber\\
 && +  \epsilon_5^{xyz} J_x^\nu J_y j_{z,\nu} + \epsilon_6^{xyz} J_x^\nu J_y^\rho j_{z,\nu\rho} + \epsilon_7^{xyz} J_x J_y^{\nu \rho} j_{z,\nu\rho} + \epsilon_8^{xyz} J_{x,\nu \alpha} J_y^{\rho \alpha} j_{z,\rho}^\nu \big]\,,
 \label{eq:short-range}
\end{eqnarray}
where $G_F=\sqrt{2}g^2/(8M^2_W)$ is the Fermi constant and $m_p$ is the proton mass. The hadronic currents are defined similarly as done in Ref.~\cite{Bergstrom:2011dt}:
\begin{equation}
 J_{R,L} = \overline{d} (1 \pm \gamma_5) u, \ \ J_{R,L}^\nu = \overline{d} \,\gamma^\nu(1 \pm \gamma_5) u,\ \ J_{R,L}^{\nu\rho} = \overline{d}\, \sigma^{\nu\rho} (1 \pm \gamma_5) u\,.
 \label{eq:hadronic-currents}
\end{equation}
The leptonic currents are defined analogously, however, for $\mu^-$-- $e^+$ conversion they must connect $\mu$-$e$ instead of $e$-$e$:
\begin{equation}
\begin{split}
  j_{R,L} &= \overline{e^c} (1 \pm \gamma_5) \mu = 2 \overline{(e_{R,L})^c}\, \mu_{R,L},\ \ j_{R,L}^\nu = \overline{e^c}\, \gamma^\nu (1 \pm \gamma_5) \mu = 2 \overline{(e_{L,R})^c} \,\gamma^\nu \mu_{R,L}\,, \\ 
  {\rm and} & \ \ j_{R,L}^{\nu\rho} = \overline{e^c}\, \sigma^{\nu\rho} (1 \pm \gamma_5) \mu=2 \overline{(e_{R,L})^c}\,\sigma^{\nu\rho} \mu_{R,L}\,.
\end{split}
 \label{eq:leptonic-currents}
\end{equation}
According to~\cite{Pas:2000vn}, the terms proportional to $\epsilon_{6,7,8}$ can be neglected for neutrinoless double beta decay. In fact, when exploiting the identity of the two electrons, they can even be shown to vanish exactly~\cite{Prezeau:2003xn} and are thus strictly inrelevant for $0\nu\beta\beta$. The same line of reasoning is not valid for $\mu^-$-- $e^+$ conversion, though, since obviously $\mu$ and $e$ are not identical.
If we restrict the discussion to the coherent part of the process ($\sim 40\%$ of all transitions~\cite{Domin:2004tk}), the outgoing positron carries away an energy of roughly $m_\mu$, while the transfer to the final state nucleus is small. In addition, one can assume that the initial and final state nuclei are non-relativistic to a good approximation. Therefore, the hadronic currents can be approximated by their non-relativistic versions, $J_\nu^-(t,\vec{x}) \simeq J_\nu^-(\vec{x}) \mathrm{e}^{i E t}$, where $E$ is the energy of the corresponding state.  By doing so, Eq.~(\ref{eq:NonRel_Hadronic}) shows that:
\begin{equation}
 \mathrm{e}^{i(E_f-E_i) t}\, J^\sigma(\vec{x}_1) J^\rho (\vec{x}_2)=\mathrm{e}^{i(E_f-E_i) t} \,J^\sigma(\vec{x}_2) J^\rho (\vec{x}_1)\,,
\end{equation}
which means that the expression is symmetric under the exchange of $\vec{x}_1 \leftrightarrow \vec{x}_2$. Given that $j_{R,L}^{\nu\rho}$ is anti-symmetric under $\rho \leftrightarrow \sigma$, the expressions related to the effective couplings $\epsilon_{6,7,8}$ will thus not contribute to the decay rate.  Note that switching to an incoherent process leads to a final state nucleus with different $J^\pi$ and an outgoing positron with reduced kinetic energy. However, as long as both initial and final state nuclei are non-relativistic and one can use a point-like vertex, the above arguments remain valid.

Treating the short-range contributions via an EFT allows for a clean separation of the nuclear physics part from the respective particle physics part, valid for a rather large class of models (namely all that realise the short-range operators under consideration). It thereby allows for a (particle-) model-independent computation of the NMEs. Consequently, it is concurrently essential to determine the relevant $\mu^-$-- $e^+$ conversion NMEs, such that limits from this LNV process can be derived~\cite{Geib:2016atx}.

\section{\label{sec:rate-explicit}Computing the decay rate: a very explicit example}

The aim of this section is to perform the computation for the decay rate for one particular short-range operator, which we choose to be $\epsilon_3^{LLR}$. This choice is motivated by several arguments:
\begin{enumerate}

\item First of all, $\epsilon_3$ is the only choice for which the NMEs have already been computed (in fact for both short- and long-range contributions~\cite{Domin:2004tk}). Ref.~\cite{Domin:2004tk} actually aimed at comparing the two cases of light and heavy neutrino exchange, with the latter realising $\epsilon_3^{LLL}$. However, once the identification with the operator coefficients has been performed (see our Sec.~\ref{sec:matching}), the results in fact carry over to our case, which in particular holds for the NME.

\item Second, while the explicit computation has been performed to some extend in Ref.~\cite{Domin:2004tk}, the computation presented mostly focuses on nuclear physics aspects and is not easily accessible for the average particle physicist. We would like to remedy this issue by presenting all relevant steps in detail, so that the pervious results are easier to use for the particle physics community.

\item After all, many aspects of the computation would not change if another operator was chosen from Eq.~\eqref{eq:short-range}. Given that all these operators are point-like, it is mainly the external projections that change, as well as the connection of the hadronic currents to the nucleus, but the more involved aspects of the computation basically remain the same.

\item We have in Ref.~\cite{Geib:2016atx} already discussed the physics potential of future experiments in constraining the operator $\epsilon_3$. Some of the results anticipated there will be much easier to grasp with the background of the explicit computation at hand. Furthermore, we have in that previous reference already listed several example models which could potentially be constrained by a future measurement of $\mu^-$-- $e^+$ conversion.

\item Finally, we will in the following also include a particular example to show how the operator $\epsilon_3$ can be obtained from a concrete underlying model. While this may sound like a slightly ambiguous strategy, it serves the additional purpose to show how the operator matching can be performed in passing when doing the full computation. The alternative, but of course equivalent, strategy would be to match on the level of Lagrangians and simply use the Feynman rules for the effective model. Since the latter option is implicitly contained in the following derivation, from Eq.~\eqref{eq:Step1} onwards, we have however decided to show an example as explicit as possible.

\end{enumerate}
Having justified our procedure, we start by quickly addressing the explicit example.

\subsection{\label{sec:rate-explicit_example}The example chosen}

\begin{figure}
\centering
 \includegraphics[width=7cm]{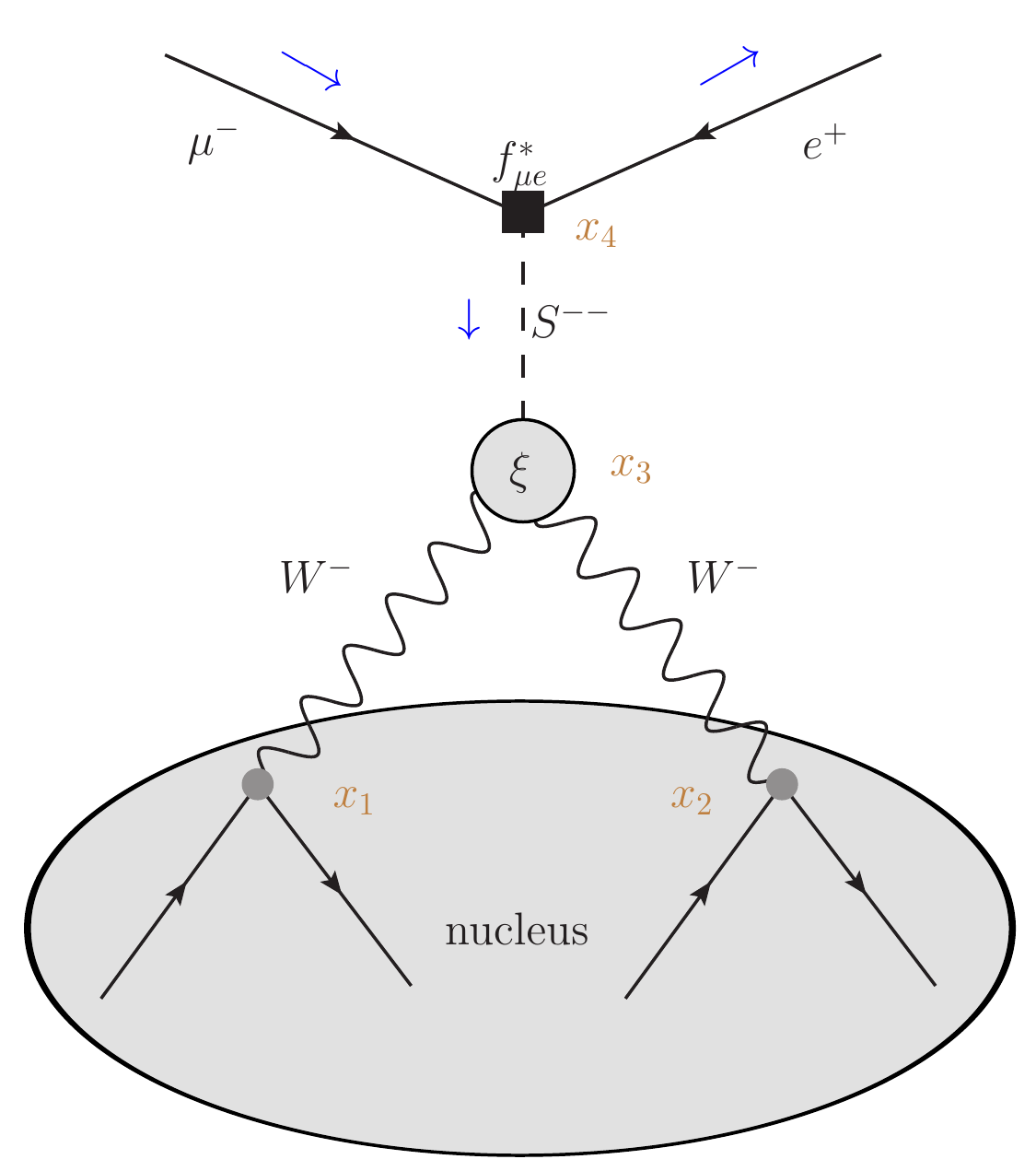}
 \caption{Realisation of $\mu^-$-- $e^+$ conversion via a doubly charged scalar. The $x_i$ denote the space-time points as used in Eq.~(\ref{eq:first_step})}
 \label{fig:DoublyCharged}
\end{figure}

While the few earlier references available~\cite{Domin:2004tk, Vergados:1980fp,Kamal:1979vw,Picciotto:1982qe,Vergados:1980em} focused on $\mu^-$-- $e^+$ conversion mediated by heavy Majorana neutrinos, we will present the computation by means of a model which extends the SM by only one doubly charged scalar~\cite{King:2014uha,Geib:2015unm,Geib:2015tvt}.  In this scenario, $\mu^-$-- $e^+$ conversion is realised via the diagram in Fig.~\ref{fig:DoublyCharged}, and the following interactions are required for its description:
\begin{equation}
 \mathcal{L}_{\text{int}}=f^*_{ab}\,S^{++}\,\overline{(l_{Ra})^c}\,l_{Rb}-\frac{g^2 v^4\,\xi}{4 \Lambda^3}\,W^+_\nu \,W^{+\nu}\,S^{--}+\frac{g}{2\sqrt{2}} V_{ud}\, W^-_\nu\, \overline{d}\,\gamma^\nu\,\big(1-\gamma_5\big)\,u\, + h.c.
 \label{eq:ex-Lan_1}
\end{equation}
Here, $\xi$ is an effective coupling connecting $S^{\pm \pm}$ to the $W$-bosons, $\Lambda$ is the UV cutoff of the model, and $f^*_{ab}=f^*_{ba}$ is symmetrically coupling the doubly charged scalar to right-handed charged leptons, see Ref.~\cite{King:2014uha} for details. Given the weak interaction contained in Eq.~\eqref{eq:ex-Lan_1}, it can already be anticipated that the only operators possibly realised from Eq.~\eqref{eq:short-range} are $\epsilon_{3,6}^{LLx}$, with $x=L,R$ (where we sloppily but conventionally refer to the operator coefficient as ``operator''). Since Fig.~\ref{fig:DoublyCharged} is realised by the effective coupling in Eq.~\eqref{eq:ex-Lan_1}, there is no operator with $\epsilon_6$. Furthermore, the doubly charged scalar $S^{--}$ coupling to the two right-handed leptons, cf.\ Eq.~\eqref{eq:ex-Lan_1}, implies that $x=R$ . Thus we expect our computation to yield a term $\epsilon_3^{LLR}$ at some point.

\subsection{\label{sec:rate-explicit_amplitude}How to obtain the amplitude}

To generate the diagram in Fig.~\ref{fig:DoublyCharged}, we need to go to fourth order in pertubation theory. Hence, the resulting leading order amplitude reads
\begin{equation}
 \langle N',f \big| S^{(4)} \big| N,i \rangle = \frac{1}{4!}\,\langle N',f \big | \int \diff^4 x_1 \diff^4 x_2 \diff^4 x_3 \diff^4 x_4\, \widehat{T}\Big\{\mathcal{L}_{\text{int}}\big(x_1\big)\mathcal{L}_{\text{int}}\big(x_2\big)\mathcal{L}_{\text{int}}\big(x_3\big)\mathcal{L}_{\text{int}}\big(x_4\big)\Big\}\big| N,i \rangle\,,
\end{equation}
where $\widehat{T}$ indicates time-ordering. The external (real) states are denoted by $\langle N' \big|$ and $\big| N \rangle$ for the final and initial state nucleus as well as $\langle f \big|$ and $\big| i \rangle$ for the final state positron and the initial bound muon. Upon assigning the space-time 4-vectors $x_i$ to fixed vertices, we obtain a combinatorical factor of $4!$. Furthermore, we need to take into account that there is an additional factor $2$ due to the symmetric property of $f^*_{ab}$, i.e., $\big[f^*_{e\mu}\overline{(l_{Re})^c}\,l_{R\mu}+f^*_{\mu e}\overline{(l_{R\mu})^c}\,l_{R e} \big]S^{++}=2\,f^*_{e\mu}\overline{(l_{Re})^c}\,l_{R\mu}\,S^{++}$. This way, the amplitude takes the form
\begin{equation}
\begin{split}
 \langle N',f \big| S^{(4)} \big| N,i \rangle =&-\frac{f^*_{e\mu}\,g^4 v^4\,\xi V_{ud}^2}{16 \Lambda^3}\langle N',f \big | \int \diff^4 x_1 \diff^4 x_2 \diff^4 x_3 \diff^4 x_4\,\widehat{T} \Big\{J_{L,\nu} \big(x_1\big) W^{-\nu}\big(x_1\big)\,J_{L,\rho} \big(x_2\big) \\
 & W^{-\rho}\big(x_2\big)\,W^+_\sigma \big(x_3\big) W^{+\sigma} \big(x_3\big) S^{--}\big(x_3\big)\,S^{++}\big(x_4\big)\overline{(l_{Re})^c}\big(x_4\big) l_{R\mu}\big(x_4\big)\Big\}\big| N,i \rangle\,.
 \end{split}
 \label{eq:first_step}
\end{equation}
The next step is to contract the boson fields. There are two indistinguishable options to contract the $W$-bosons, which leads to an additional factor $2$:
\begin{equation}
\begin{split}
 &\widehat{T}\Big\{W^{-\nu}\big(x_1\big)\,W^{-\rho}\big(x_2\big)\,W^+_\sigma \big(x_3\big) W^{+\sigma} \big(x_3\big)\, S^{--}\big(x_3\big)\,S^{++}\big(x_4\big)\Big\}\\
 &=\widehat{N}\Big\{
 \contraction{}{W}{{}^{-\nu}\big(x_1\big)\,W^{-\rho}\big(x_2\big)\,}{W}
 \contraction[2ex]{W^{-\nu}\big(x_1\big)\,}{W}{{}^{-\rho}\big(x_2\big)\,W^+_\sigma \big(x_3\big)}{W}
 W^{-\nu}\big(x_1\big)\,W^{-\rho}\big(x_2\big)\,W^+_\sigma \big(x_3\big) W^{+\sigma} \big(x_3\big)\, S^{--}\big(x_3\big)\,S^{++}\big(x_4\big)
 \\
 &+
 \contraction[2ex]{}{W}{{}^{-\nu}\big(x_1\big)\,W^{-\rho}\big(x_2\big)\,W^+_\sigma \big(x_3\big)}{W}
 \contraction{W^{-\nu}\big(x_1\big)\,}{W}{{}^{-\rho}\big(x_2\big)\,}{W}
 W^{-\nu}\big(x_1\big)\,W^{-\rho}\big(x_2\big)\,W^+_\sigma \big(x_3\big) W^{+\sigma} \big(x_3\big)\, S^{--}\big(x_3\big)\,S^{++}\big(x_4\big)
 \Big\}\\
 =&\, 2\, \Delta_W^{\nu \sigma}\big(x_1-x_3\big)\Delta_{W\,\sigma}^{\rho} \big(x_2-x_3\big)\,\Delta_S \big(x_3-x_4\big)\,,
 \end{split}
\end{equation}
when phrased in terms of propagators in coordinate space. The propagators in coordinate space can be written as Fourier transforms of their momentum space representations,
\begin{equation}
 \begin{split}
  \Delta_S \big(x-y\big)=& \int \frac{\diff^d q}{(2\pi)^d}\, \mathrm{e}^{-i q\cdot (x-y)}\,\frac{i}{q^2-M^2_S+i\varepsilon}\,,\\
  \Delta_W^{\nu \sigma}\big(x-y\big)=& \int \frac{\diff^d q}{(2\pi)^d}\, \mathrm{e}^{-i q\cdot (x-y)}\,\frac{-i}{q^2-M^2_W+i\varepsilon}\big( g^{\nu \sigma}-\frac{q^\nu q^\sigma}{M^2_W}\big)\,,
 \end{split}
\end{equation}
where the momenta $q$ propagate from space-time point $y$ to $x$, respectively. That way, we introduce the momenta $q$ (propagating from $x_4$ to $x_3$), $l$ (propagating from $x_3$ to $x_2$), and $k$ (propagating from $x_3$ to $x_1$).

At this point, instead of obstinately pursuing the computation, it is useful to take a closer look at the energy scales of the conversion process, see e.g.~\cite{Simkovic:2000ma,Domin:2004tk}. We only consider the \emph{coherent} process, which means that both the initial and final nucleus are in the ground state. Although the coherent process is estimated to constitute only about $\sim 40 \%$ of the total amount of nuclear transitions~\cite{Divari:2002sq}, it is experimentally favoured due to its minimal background for the outgoing positron, as it carries away the maximal energy. The basic concept of the $\mu^-$-- $e^+$ conversion is that a muon is trapped by an atom, cascades down in energy levels until it is bound in the $1s$ state, and gets then captured by the nucleus, thereby emitting a positron. The total energy of the muon in the $1s$ bound state is given by $E_\mu=m_\mu-\varepsilon_b$, where $m_\mu$ is the muon mass and $\varepsilon_b$ is its binding enery. Since the binding energy is roughly $\varepsilon_b \simeq \frac{m_\mu }{m_e} \cdot 13.6~\mathrm{eV}\cdot Z \ll m_\mu$, the muon can in any case be considered as non-relativistic. The energy of the positron hence results into:
\begin{equation}
 E_e=\underbrace{m_\mu-\varepsilon_b}_{\sim \mathcal{O}(100~\mathrm{MeV})}-\underbrace{(E_f-E_i)}_{\sim \mathcal{O}(\mathrm{MeV})} \sim \mathcal{O}(100~\mathrm{MeV})\,,
\end{equation}
where $E_{i,f}$ are the energies of the initial and final nuclear ground states, respectively. Both nuclei are -- to a good approximation -- at rest, which in combination with the nuclei not being excited leads to $E_f-E_i\sim \mathcal{O}(\mathrm{MeV})$~\cite{Kamal:1979vw}.

Two things can therefore be concluded:
\begin{enumerate}

\item The positron energy peaks around $m_\mu$, which allows for a clear separation from possible background positrons stemming from, e.g., $\beta^+$ decay stemming from potential impurities. This will hold as long as the experiment is able to distinguish positrons from electrons (which is non-trivial if they are fast).

\item The energy transfer from the bound muon to the nucleus is small, $\mathcal{O}(\mathrm{MeV})$, which implies that $l^2,k^2,q^2 \ll M^2_S,M^2_W$. The latter amounts to effectively integrating out both the $W$-bosons and the doubly charged scalar.

\end{enumerate}

Upon contracting the bosonic propagators, the matrix element takes the form
\begin{equation}
\begin{split}
 -i\,&\frac{f^*_{e\mu}\,g^4 v^4\,\xi V_{ud}^2}{8 \Lambda^3\,M_S^2\,M^4_W}\int \diff^4 x_1 \diff^4 x_2 \diff^4 x_3 \diff^4 x_4 \int \frac{\diff^4 q \diff^4 k \diff^4 l}{(2\pi)^{12}}\,\langle N',f \big |\,\mathrm{e}^{-iq\cdot(x_3-x_4)}\mathrm{e}^{-ik\cdot(x_1-x_3)}\\
 & \mathrm{e}^{-il\cdot(x_2-x_3)}\,\widehat{T} \Big\{J_{L,\nu} \big(x_1\big) J_L^\nu \big(x_2 \big) \underbrace{\overline{(l_{Re})^c}\big(x_4\big) l_{R\mu}\big(x_4\big) }_{\equiv j_R (x_4)} \Big\} \big| N,i \rangle\,.
 \end{split}
\end{equation}
Since we contracted the gauge boson propagators, it is reasonable to also switch to a notation using Fermi's constant, i.e., $G_F/\sqrt{2}=g^2/(8 M^2_W)$. At this point, we can also identify the short-range operator coefficient. For brevity, we introduce the operator coefficient that is realised in this scenario, as derived in more detail in Sec.~\ref{sec:matching_general}:
\begin{equation}
 \epsilon^{LLR}_3 \equiv 4 V^2_{ud}\,m_p\,\frac{f^*_{e\mu} v^4\,\xi}{\Lambda^3 M^2_S}.
 \label{eq:example-identification}
\end{equation}
We furthermore note that $x_3$-dependences solely remain in the exponential functions. Hence, we obtain a four-dimensional delta-function, $(2\pi)^4\,\delta^{(4)}(l-q+k)$, upon performing the $x_3$-integration. We can dispose of the $l$-integration subsequently. That way, we obtain:
\begin{eqnarray}
 && -i\, \frac{ G_F^2}{m_p}\,\epsilon^{LLR}_3\int \diff^4 x_1 \diff^4 x_2 \diff^4 x_4 \int \frac{\diff^4 q \diff^4 k}{(2\pi)^{8}}\,\nonumber\\
 && \langle N',f \big |\,\mathrm{e}^{iq\cdot x_4}\mathrm{e}^{-ik\cdot x_1}\mathrm{e}^{-i (q-k)\cdot x_2} \widehat{T} \Big\{J_{L,\nu} \big(x_1\big) J_L^\nu \big(x_2 \big) j_R (x_4) \Big\} \big| N,i \rangle\,.
 \label{eq:Step1}
\end{eqnarray}
Next, we consider the remaining structures, 
\begin{equation}
 \langle N',f \big |\widehat{T} \Big\{J_{L,\nu} \big(x_1\big) J_L^\nu \big(x_2 \big) j_R (x_4) \Big\} \big| N,i \rangle = \langle N' \big |\widehat{T} \Big\{J_{L,\nu} \big(x_1\big) J_L^\nu \big(x_2 \big) \Big\} \big| N \rangle\langle f \left| j_R (x_4) \right| i \rangle \,,
\end{equation}
which allows to split the structure into hadronic and leptonic parts.

Starting with the leptonic part, we need to take into account that neither the muon nor the positron are freely propagating. The muon is bound in the $1s$ state, whereas the positron is a free particle which propagates under the influence of the Coulomb potential of the nucleus. Consequently, we need to modify the spinors $u$ and $v$ of the muon and the positron, respectively, to describe a bound state and a continuum state subject to a potential, instead of freely propagating particles and antiparticles. This can be done by using~\cite{Kamal:1979vw,Domin:2004tk}\footnote{Note that \cite{Domin:2004tk} uses another normalisation for the spinors than we do, which also reflects in different relations for the spin sums. The translation will be discussed in Appendix~\ref{app:A}.}
\begin{equation}
 \begin{split}
 u^{\text{free}}_\mu &\rightarrow \phi_\mu (\vec{x}_4)\,u^{\text{free}}_\mu\,\\
 \text{and} \quad v^{\text{free}}_e &\rightarrow \sqrt{F(Z-2,E_e)}\, v^{\text{free}}_e\,,
 \end{split}
\end{equation}
where the bound muon wave function $\phi_\mu$ and the Fermi function $F(Z,E)$ are given by\footnote{Note that we consistently use the non-relativistic approximation for the bound muon wave function. Note also that the sign of $y$ is opposite to the the usual one quoted in Fermi functions, due to the emitted particle being a positron rather than an electron, cf.\ Appendix F.3 in Ref.~\cite{Doi:1985dx}.}
\begin{equation*}
 \phi_\mu (\vec{x})= \frac{Z^{3/2}}{\big(\pi a^3_\mu \big)^{1/2}}\,\mathrm{e}^{-\frac{Z}{a_\mu}|\vec{x}|} \quad \text{and} \quad F(Z,E)=\Bigg[\frac{2}{\Gamma\big[2\gamma_1+1\big]}\Bigg]^2 \big(2 |\vec{p}_e| R\big)^{2(\gamma_1-1)} \big|\Gamma\big[\gamma_1-i y \big] \big|^2 \mathrm{e}^{-\pi y}\,.
\end{equation*}
Here, $a_\mu=4\pi/(m_\mu e^2)$ is the muon's Bohr radius, $\gamma_1=\sqrt{1-(\alpha Z)^2}$, $y=\alpha Z E/|\vec{p}_e|$, and $\alpha \simeq 1/137$ is the fine structure constant. Furthermore, $Z$ denotes the atomic number and $R=1.1\,A^{1/3}~\mathrm{fm}$ the nuclear radius for an atom with mass number $A$. We will abbreviate $u_\mu \equiv u^{\text{free}}_\mu$ and $v_e \equiv v^{\text{free}}_e$ in the following. That way, the leptonic part of the amplitude can be rewritten such that
\begin{equation}
 \langle f | j_R (x_4) | i \rangle = \mathrm{e}^{i k_e \cdot x_4} \underbrace{\mathrm{e}^{-i k_\mu \cdot x_4}}_{\approx \mathrm{e}^{-i E_\mu \cdot x^0_4}} \sqrt{F(Z-2,E_e)}\,\phi_\mu (\vec{x}_4)\,\overline{v_e}(k_e)\,\mathrm{P_R}\,u_\mu (k_\mu)\,,
 \label{eq:Leptonic1}
\end{equation}
with $\mathrm{P_R}\equiv (1+\gamma_5)/2$ and the muon (positron) momentum denoted by $k_\mu$ ($k_e$). Given that we assume the muon to be non-relativistic, we can thus simplify Eq.~\eqref{eq:Step1} using
\begin{equation}
\begin{split}
 &\int \diff^4 x_{1,2,4} \int \frac{\diff^4 q \diff^4 k}{(2\pi)^{8}}\,\mathrm{e}^{iq\cdot x_4}\mathrm{e}^{-ik\cdot x_1}\mathrm{e}^{-i (q-k)\cdot x_2}\mathrm{e}^{i k_e \cdot x_4}\mathrm{e}^{-i E_\mu \cdot x^0_4}\\
 =& \int \diff^4 x_{1,2} \diff^3 x_4 \int \frac{\diff^3 q \diff^4 k}{(2\pi)^{7}}\,\mathrm{e}^{-i k\cdot (x_1-x_2)}\mathrm{e}^{-i \vec{q}\cdot (\vec{x}_4-\vec{x}_2)}\mathrm{e}^{-i \vec{k}_e \cdot \vec{x}_4}\mathrm{e}^{-i(E_\mu-E_e)x_2^0}\,.
 \end{split}
\end{equation}
Moving on to the hadronic part, we need to incorporate the information that the quarks are not locally fixed, but instead distributed within the nucleons. This can be done by introducing so-called nucleon form factors, which model the charge distribution. We use the dipole parametrisation such that
\begin{equation}
 \tilde{F} (\vec{p}\,^2,\,\Lambda_i)=\frac{1}{\big(1+\vec{p}\,^2/\Lambda_i^2\big)^2}\,,
\end{equation}
where the scale $\Lambda_i\sim \mathcal{O}(\mathrm{GeV})$ depends on how the quarks interact. As there are two nucleon interactions taking place, we thus include an additional factor of $\tilde{F}(\vec{k}\,^2,\,\Lambda_i)\,\tilde{F}((\vec{k}-\vec{q}\,)^2,\,\Lambda_i)$. We can neglect the $\vec{q}$-dependence due to the momentum transfer being of the order $m_\mu \ll \Lambda_i$. As a result, not only the $k^0$- but also the $\vec{q}$-dependence restricts itself to the exponential functions, allowing for
\begin{equation}
\begin{split}
 &\int \diff^4 x_{1,2} \diff^3 x_4 \int \frac{\diff^3 q \diff^4 k}{(2\pi)^{7}}\,\mathrm{e}^{-i k\cdot (x_1-x_2)}\mathrm{e}^{-i \vec{q}\cdot (\vec{x}_4-\vec{x}_2)}\mathrm{e}^{-i \vec{k}_e \cdot \vec{x}_4}\mathrm{e}^{-i(E_\mu-E_e)x_2^0}\,\tilde{F}^2(\vec{k}\,^2,\,\Lambda_i)\,\phi_\mu (\vec{x}_4)\\
 =& \int \diff^4 x_{1,2} \int \frac{\diff^3 k}{(2\pi)^{3}}\,\mathrm{e}^{i \vec{k}\cdot (\vec{x}_1-\vec{x}_2)}\mathrm{e}^{-i \vec{k}_e \cdot \vec{x}_2}\mathrm{e}^{-i(E_\mu-E_e)x_2^0}\,\delta(x_1^0-x_2^0)\,\tilde{F}^2(\vec{k}\,^2,\,\Lambda_i)\,\phi_\mu (\vec{x}_2)\,.
 \end{split}
\end{equation}
Moreover, we can re-express the hadronic part using a non-relativistic approximation, which leads to
\begin{equation}
\begin{split}
 &\,\langle N' \big |\widehat{T} \Big\{J_{L,\nu} \big(x_1\big) J_L^\nu \big(x_2 \big) \Big\} \big| N \rangle= \sum_n \Big\{ \Theta(x_1^0-x_2^0)\,\mathrm{e}^{i(E_f-E_n)x_1^0}\,\mathrm{e}^{i(E_n-E_i)x_2^0}\langle N'\big| J_{L,\nu}(\vec{x}_1)\big| n\rangle\\
 &\langle n \big| J_L^\nu (\vec{x}_2) \big| N \rangle +\Theta(x_2^0-x_1^0)\,\mathrm{e}^{i(E_f-E_n)x_2^0}\,\mathrm{e}^{i(E_n-E_i)x_1^0}\langle N'\big| J_{L,\nu}(\vec{x}_2)\big| n\rangle
 \langle n \big| J_L^\nu (\vec{x}_1) \big| N \rangle \Big\}\,,
 \end{split}
 \label{eq:NonRel_Hadronic}
\end{equation}
where $\Theta(x)$ denotes the Heaviside function. Here, we take the sum over the virtual intermediate nuclear states labeled by $n$ and make use of $J_{L,\nu} (\vec{x})\equiv J_{L,\nu} (0,\vec{x})$. Further simplifications arise from the aforementioned considerations that implied $x^0_1=x^0_2$. The latter results into $n$-independent factors which allow us to carry out the sum explicitly and make use of the completeness of the set of states introduced: $\sum\nolimits_n |n \rangle \langle n|=1$. In combination with $\Theta (0)=1/2$, the hadronic part, Eq.~\eqref{eq:NonRel_Hadronic}, takes the following form after performing the $x^0_1$-integration:
\begin{equation}
 \mathrm{e}^{i(E_f-E_i)x_2^0} \langle N' \big |J_\nu^-(\vec{x}_1) J^{-\nu}(\vec{x}_2) \big| N \rangle\,.
\end{equation}
Checking for $x_2^0$-dependences we note that, at this point, $x^0_2$ only appears in exponents. Upon carrying out this integration, we finally obtain the conservation of external energies, $(2\pi)\,\delta(E_f-E_i+E_e-E_\mu)$, as to be expected.

Combining these modifications that enter due to the physical properties of the process, we can rewrite Eq.~\eqref{eq:Step1} into 
\begin{equation}
\begin{split}
-i\,& \frac{G_F^2\epsilon^{LLR}_3}{m_p}\,(2\pi)\,\delta(E_f-E_i+E_e-E_\mu)\,\overline{v_e}(k_e)\,\mathrm{P_R}\,u_\mu (k_\mu)\,\int \diff^3 x_{1,2}\int \frac{\diff^3 k}{(2\pi)^{3}}\\
& \langle N' \big |\,\mathrm{e}^{i \vec{k}\cdot (\vec{x}_1-\vec{x}_2)}\mathrm{e}^{-i \vec{k}_e \cdot \vec{x}_2}\,\tilde{F}^2(\vec{k}\,^2,\,\Lambda_i)\,\sqrt{F(Z-2,E_e)}\,\phi_\mu (\vec{x}_2)\,J_{L,\nu}(\vec{x}_1) J_L^\nu(\vec{x}_2)\big| N \rangle\,.
 \end{split}
 \label{eq:Step2}
\end{equation}
Within the non-relativistic approximation, the hadronic currents can be written in terms of effective transition operators. These consist of the basic spin and isospin structures. In principle, there are five spin structures and only one isospin structure~\cite{Divari:2002sq}. As explained in~\cite{Divari:2002sq,Vergados:1980fp,Domin:2004tk,Simkovic:2000ma}, however, two spin structures are expected to be most important: the Fermi ($\propto g_V$) and Gamow-Teller ($\propto g_A$) parts. Hence, the hadronic current can be rephrased as
\begin{equation}
 \tilde{F}(\vec{k}\,^2,\,\Lambda_i)\,J^-_\nu (\vec{x}\,)\to \sum_m \tau^-_m\,\left(g_V\, \tilde{F}(\vec{k}\,^2,\,\Lambda_V)\,g_{\nu 0} +g_A\,\tilde{F}(\vec{k}\,^2,\,\Lambda_A)\,g_{\nu j}\,\sigma^{j}_m \right)\,\delta^{(3)}(\vec{x}-\vec{r}_m\,)\,,
 \label{eq:FGT}
\end{equation}
where we sum over all nucleons, with $\vec{r}_m$ being the position of the $m$-th nucleon. Here, $\tau^-_m$ is the nuclear isospin raising operator, which means that it can change protons into neutrons (as needed for $\mu^-$-- $e^+$ conversion):
\begin{equation}
 \tau^-_m \,\big|\text{proton} \rangle_m= \big| \text{neutron} \rangle_m \quad \text{and} \quad \tau^-_m \,\big|\text{neutron} \rangle_m= 0\,.
\end{equation}
The Gamow-Teller operator flips the spin of the $m$-th nucleon into the $j$-th direction. Note that we have employed different scales $\Lambda_i$, with $i = V, A$, depending on the type of interaction. Generic values are $\Lambda_V=0.71~\mathrm{GeV}$ and $\Lambda_A=1.09~\mathrm{GeV}$~\cite{Domin:2004tk}.

Now we have collected all ingredients to obtain the final version of the amplitude. Using Eq.~\eqref{eq:Step2} together with
\begin{eqnarray}
 &&\tilde{F}(\vec{k}\,^2,\,\Lambda_{i_1})\,J_{L,\nu} (\vec{x}_1)\,\tilde{F}(\vec{k}\,^2,\,\Lambda_{i_2})\,J_L^\nu (\vec{x}_2) \\
 &&\to \sum_{m,l} \tau^-_m \tau^-_l\,\Big(g^2_V\, \tilde{F}^2(\vec{k}\,^2,\,\Lambda_V)-g^2_A\,\tilde{F}^2(\vec{k}\,^2,\,\Lambda_A)\,\vec{\sigma}_m\cdot \vec{\sigma}_l\Big) \,\delta^{(3)}(\vec{x}_1-\vec{r}_m\,)\,\delta^{(3)}(\vec{x}_2-\vec{r}_l\,)\,, \nonumber
\end{eqnarray}
we obtain the final version of the amplitude:
\begin{equation}
 \mathcal{M}= \frac{G_F^2 \epsilon^{LLR}_3 g_A^2 m_e}{2 R}\,\sqrt{F(Z-2,E_e)}\,\delta(E_f-E_i+E_e-E_\mu)\,\overline{v_e}(k_e)\,\mathrm{P_R}\,u_\mu (k_\mu)\,\mathcal{M}^{(\mu^-,e^+)\,\phi}\,.
\label{eq:Amplitude1}
\end{equation}
In accordance with Ref.~\cite{Domin:2004tk}, cf.\ App.~\ref{app:B} in order to understand the equivalence in detail, we define {\it nuclear matrix element} to be\footnote{Note that Eqs.~(37) and~(49) in Ref.~\cite{Domin:2004tk}, which both are supposed to contain expressions for the NME in case of a realisation via heavy Majorana neutrinos, differ by a factor of~2. After carefully checking an analogous discussion for $0\nu\beta \beta$~\cite{Simkovic:1999re}, we reckon that Eq.~(49) of~\cite{Simkovic:1999re} is the correct normalisation, while the additional factor of~2 in Eq.~(37) of~\cite{Simkovic:1999re} is a typo. Our matrix element in Eq.~\eqref{eq:NME1} is defined to be consistent with Eq.~(49) in~\cite{Domin:2004tk}.}
\begin{eqnarray}
 \mathcal{M}^{(\mu^-,e^+)\,\phi} &\equiv& \frac{4\pi}{(2\pi)^{3}}\,\frac{R}{m_p m_e}\,\int \diff^3 k\,\langle N' \big |\,\sum_{m,l} \tau^-_m \tau^-_l\,\bigg[\tilde{F}^2\big(\vec{k}\,^2,\,\Lambda_A\big)\,\vec{\sigma}_m\cdot \vec{\sigma}_l-\frac{g_V^2}{g_A^2}\,\tilde{F}^2\big(\vec{k}\,^2,\,\Lambda_V\big)\bigg]\nonumber\\
 &&\mathrm{e}^{i \vec{k}\cdot (\vec{r}_m-\vec{r}_l)}\mathrm{e}^{-i \vec{k}_e \cdot \vec{r}_l}\,\phi_\mu (\vec{r}_l)\,\big| N \rangle\,.
\label{eq:NME1}
\end{eqnarray}
We have now reached an important point: once the reader's favourite nuclear physics expert has computed a numerical values for the NME $\mathcal{M}^{(\mu^-,e^+)\,\phi}$, this can be directly inserted into Eq.~\eqref{eq:Amplitude1} and used to constrain any particle physics model leading to the operator $\epsilon^{LLR}_3$. The same could in principle be done for all other short-range operators in Eq.~\eqref{eq:short-range}, provided that the corresponding NMEs are known. 

\subsection{\label{sec:rate-explicit_rate}From the amplitude to the decay rate}

In order to derive the decay rate from the matrix element obtained above, we need to employ Fermi's Golden Rule,
\begin{equation}
 \Gamma = 2\pi \, \frac{V/T}{(2\pi)^3} \int \diff^3 k_e \, \big|\mathcal{M}\big|^2\,,
\end{equation}
where an integral over the positron's momentum $k_e$ is performed. Here, $T$ is some time interval covering the process, and $V$ some volume that we set to be unity, i.e., $V=1$. The latter was already done silently when introducing the electron wave function in Eq.~\eqref{eq:Leptonic1}.\footnote{In our normalisation with respect to free spinors, the wave function is given by $\overline{\psi}_e = \mathrm{e}^{i k_e\cdot x} \overline{v}_e$, where $V\equiv 1$ has already been employed. So, the electron wave function is 'normalised to one particle in the volume $V=1$'~\cite{Berestetsky:1982aq}.}

Next, we take the spin average over the initial and the spin sum over the final states. With respect to the free spinors and the Lorentz structure, we deal with the expression
\begin{equation}
 \frac{1}{2} \sum_{r,s} \big|\overline{v_e^r} (k_e)\,\mathrm{P_R}\, u^s_\mu (k_\mu) \big|^2=\frac{1}{4}\,.
\end{equation}
Note that we obtain this result independently of the normalisation that was used for the free spinors, because the normalisation of the spinors and the according spin sums ultimately cancel in the squared amplitude (as they should in order to yield a consistent result). In doing so, we get
\begin{equation}
 \frac{1}{2}\sum_{\text{spins}}\big|i \mathcal{M}\big|^2 = \frac{g_A^4 m^2_e G_F^4 |\epsilon^{LLR}_3|^2}{16R^2}\, \big|F(Z-2,E_e)\big| \,\text{``}\delta(E_f-E_i+E_e-E_\mu)^2\text{''}\,\big|\mathcal{M}^{(\mu^-,e^+)\,\phi}\big|^2\,,
\end{equation}
where we encounter two issues that we need to discuss briefly.

First of all, we can assume to good approximation that the muon wave function only varies slowly within the nucleus, which is justified both by the muon being non-relativistic and by the size of the nucleus being tiny compared to the muon's Bohr radius. Thus, the following standard approximation is valid~\cite{Kosmas:1993ch}:
\begin{equation}
 \big|\mathcal{M}^{(\mu^-,e^+)\,\phi}\big|^2=\langle \phi_\mu \rangle^2\,\big|\mathcal{M}^{(\mu^-,e^+)}\big|^2\quad \text{with} \quad \big|\mathcal{M}^{(\mu^-,e^+)}\big|=\big|\mathcal{M}^{(\mu^-,e^+)\,\phi}\big|_{\phi=1}\,.
\end{equation}
We can use $\langle \phi_\mu \rangle^2=\frac{\alpha^3 m^3_\mu}{\pi}\frac{Z_{\text{eff}}}{Z}$ as an approximation for the muon average probability density~\cite{Kosmas:1993ch}, where $Z_{\text{eff}}$ denotes the effective atomic charge that accounts for the deviation from the wave function at the origin~\cite{Simkovic:2000ma}. It can be conveniently obtained by taking the average of the muon wave function over the nuclear density~\cite{Kuno:1999jp,Chiang:1993xz}.

Second, we also encounter the standard ``issue'' of squaring the delta function. How to treat this square is discussed thoroughly in many textbooks, see e.g.\ Ref.~\cite{Bjorken:1964}, and it results in 
\begin{equation}
 \text{``}\delta(E_f-E_i+E_e-E_\mu)^2\text{''}=\frac{T}{2\pi}\,\delta(E_f-E_i+E_e-E_\mu)\,.
\end{equation}

Putting everything together, the decay rate takes the form
\begin{equation}
 \Gamma = \frac{g_A^4 m^2_e G_F^4 |\epsilon^{LLR}_3|^2}{16R^2}\, \big|F(Z-2,E_e)\big| \,\langle \phi_\mu \rangle^2\,\int \frac{\diff^3 k_e}{(2\pi)^3}\delta(E_f-E_i+E_e-E_\mu)\,\big|\mathcal{M}^{(\mu^-,e^+)}\big|^2\,.
\end{equation}
For a coherent transition, we can assume that $E_i\simeq E_f$. In addition, we take the positron to be highly relativistic, i.e.\ $E_e\simeq |\vec{k}_e|$, while the muon is perfectly non-relativistic, i.e.\ $E_\mu\simeq m_\mu$. As a consequence, the delta function reduces considerably, $\delta(E_f-E_i+E_e-E_\mu) \to \delta(|\vec{k}_e|-m_\mu)$. Furthermore, as shown in App.~\ref{app:B}, the NME only depends on the absolute value of $\vec{k}_e$ but not on its direction. Hence, the angular integration simply provides a factor of $4\pi$, and the remaining $|\vec{k}_e|$ integration only enforces $|\vec{k}_e|=m_\mu$. So, the decay rate -- after performing the phase space integration of the NME -- takes its final form:
\begin{equation}
 \Gamma = \frac{g^4_A\, G_F^4\, m^2_e\, m^2_\mu\, |\epsilon^{LLR}_3|^2}{32\pi^2 R^2}\, \big|F(Z-2,E_e)\big| \,\langle \phi_\mu \rangle^2\,\big|\mathcal{M}^{(\mu^-,e^+)}\big|^2\,.
 \label{eq:final-rate}
\end{equation}
The NME used here is obtained from combining Eqs.~\eqref{eq:NME1} and~\eqref{eq:AngularFinal}:
\begin{equation}
\begin{split}
 \mathcal{M}^{(\mu^-,e^+)}= &\frac{8 R}{m_p m_e}\,\int \diff k \,k^2\,\langle N' \big |\,\sum_{m,l} \tau^-_m \tau^-_l\,\bigg[\tilde{F}^2\big(\vec{k}\,^2,\,\Lambda_A\big)\,\vec{\sigma}_m\cdot \vec{\sigma}_l-\frac{g_V^2}{g_A^2}\,\tilde{F}^2\big(\vec{k}\,^2,\,\Lambda_V\big)\bigg]\\
&j_0 \big(k r_{lm}\big)\,\sum_{\lambda}\,\sqrt{2\lambda +1}\, j_{\lambda}\big(k_e R_{lm}\big)j_{\lambda}\big(k_e r_{lm}/2\big)\,\Big\{ Y_{\lambda} \big(\Omega_{r_{lm}} \big)\otimes Y_{\lambda} \big(\Omega_{R_{lm}}\big)\Big\}_{00}\,\big| N \rangle\,.
\end{split}
\end{equation}
Note that our decay rate, Eq.~\eqref{eq:final-rate}, differs from the one obtained in~\cite{Domin:2004tk} by a factor of $\pi$, even upon using the translations discussed in App.~\ref{app:A}. The tension between the results only appears at the level of decay rates. As shown in App.~\ref{app:A}, the results agree on the level of amplitudes.

Eq.~\eqref{eq:final-rate} is the desired result: given a concrete particle physics model that reproduces $\epsilon_3$, one can match this operator coefficient to fundamental model parameters. As soon as the NME $\mathcal{M}^{(\mu^-,e^+)}$ is known, all other quantities contained in the decay rate are either known constants of  Nature\footnote{Note that, however, the value of the axial vector coupling $g_A$ may be affected by \emph{quenching}~\cite{Barea:2013bz}, similar to $0\nu\beta\beta$.} or they can be computed easily. Apart from the obvious dependence on the NME, nuclear characteristics are contained in the radius $R$ (i.e., the atomic number $A$) and in the Fermi function $F(Z-2,E_e)$. However, at least for the set of isotopes discussed in the literature on muon conversion, the main variation with $Z$ and/or $A$ lies within the NME itself, whereas all other isotope-dependent quantities vary comparativley mildly.

\section{\label{sec:matching}Matching concrete particle physics models onto effective operator coefficients}

In this section, we will discuss how to map certain particle physics models onto the effective operator coefficients contained in Eq.~\eqref{eq:short-range}. Given that in our  computation performed in the previous section we have drawn the explicit comparison to Ref.~\cite{Domin:2004tk} at several places, we will start this section by a simplified discussion focusing on drawing the parallels between the heavy neutrino exchange discussed in that reference and our example model featuring the doubly charged scalar. We will then present a more detailed discussion on how to obtain $\epsilon$-coefficients from several concrete models. Feynman rules which may be necessary to reproduce our results are listed in App.~\ref{app:C}.

\subsection{\label{sec:matching_simple}Heavy neutrino exchange vs.\ doubly charged scalar exchange}

In order to use the NMEs as derived for the exchange of heavy Majorana neutrinos~\cite{Domin:2004tk}, we calculated the conversion amplitude, factorised it into particle and nuclear physics contributions, and determined the factorised decay rate, see Sec.~\ref{sec:rate-explicit}. To further check if we performed every step of the computation consistently, we now match the amputated diagram for the realisation of $\mu^-$-- $e^+$ conversion via the doubly charged scalar to the version with the heavy Majorana neutrino. With this procedure we can compare our decay rate, Eq.~\eqref{eq:final-rate} where the short-range operator was explicitly given by Eq.~\eqref{eq:example-identification}, with Eq.~(50) from~\cite{Domin:2004tk}. 
\begin{figure}[h]
 \centering
 \begin{minipage}{4.5cm}
  \includegraphics[width=4.5cm]{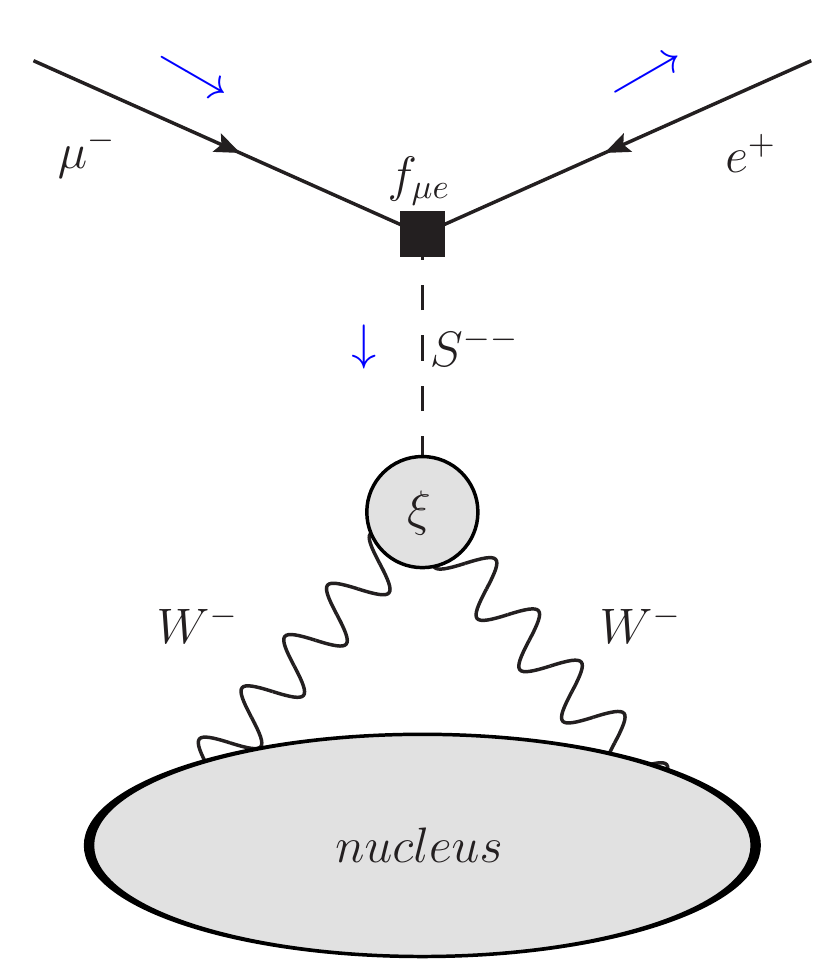}
 \end{minipage}
 \hspace{0.4cm}
 $\Longleftrightarrow$
 \hspace{0.4cm}
 \begin{minipage}{7cm}
  \includegraphics[width=7cm]{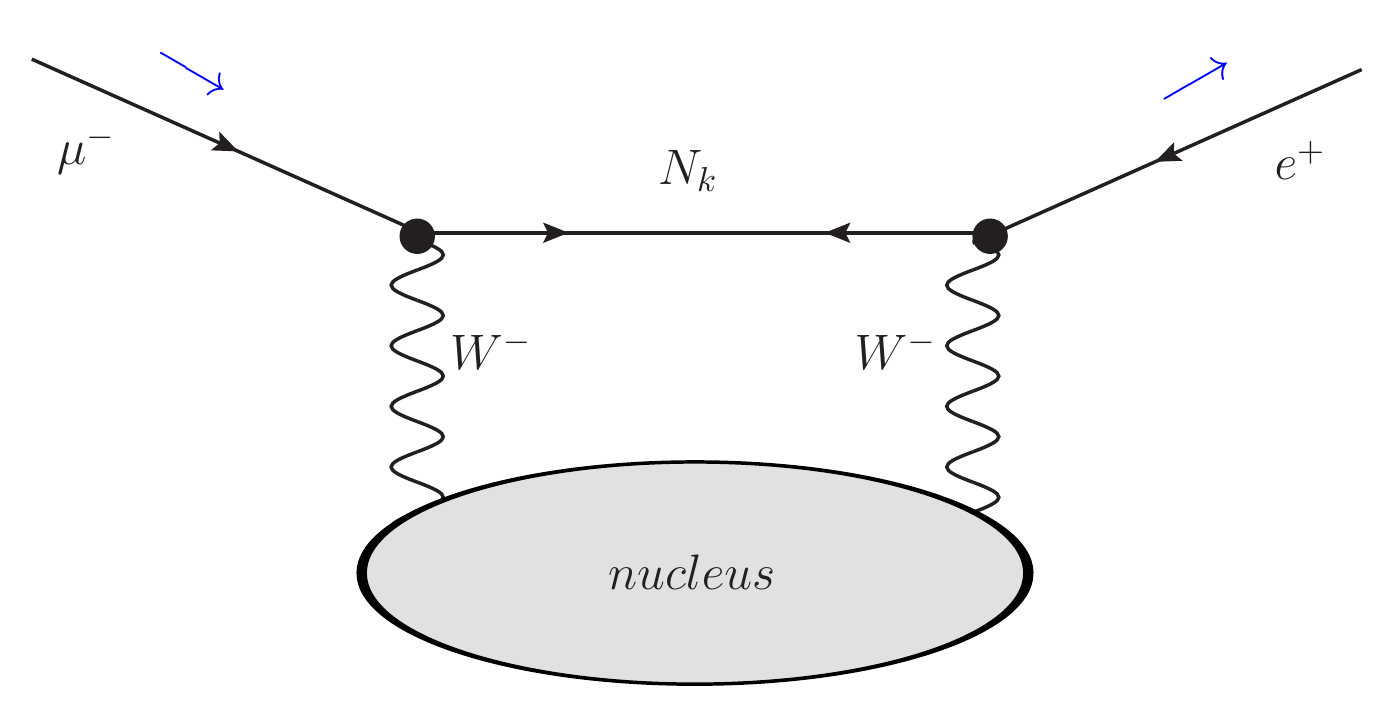}
 \end{minipage}
\caption{\label{Fig:CompareModels}Realisation of $\mu^-$-$e^+$ conversion via (left): a doubly charged scalar or (right): a heavy Majorana neutrino.}
\end{figure}

From the Feynman diagrams in Fig.~\ref{Fig:CompareModels}, we obtain two amplitudes from which we amputate the nuclear parts. This is a reasonable procedure since for both realisations of the $\mu^-$-- $e^+$ conversion the nuclear part of the process -- depicted by the greyish circle -- is identical. Starting with the left-hand side of Fig.~\ref{Fig:CompareModels}, we extract
\begin{eqnarray}
 &&\overline{v}_e\, \mathrm{P_R}\, u_\mu \, \frac{-i g^2\,f^*_{e\mu} \, v^4 \,\xi\,g_{\sigma' \rho'}}{\Lambda^3}\,\frac{i}{q^2-M^2_S}\, \frac{-i g^{\sigma \sigma '}}{k_1^2-M^2_W}\, \frac{-i g^{\rho \rho '}}{k_2^2-M^2_W}\nonumber\\
 &&\xrightarrow{M^2_{S,W} \gg k^2_{1,2},q^2}\quad \overline{v}_e\, \mathrm{P_R}\, u_\mu \, \frac{i g^2\, f^*_{e\mu} \,v^4 \,\xi\,g^{\sigma  \rho }}{M^4_W\,\Lambda^3\, M^2_S}\,.
\label{eq:DoublyAmputated}
\end{eqnarray}
Analogously, we obtain 
\begin{eqnarray}
 &&\sum_k\,\overline{v_e}\,\gamma_{\rho '}\,\mathrm{P_R}\,\gamma_{\sigma '}\,u_\mu \,V_{ek} V_{\mu k}\,\frac{g^2}{2}\,\frac{i M_k}{l^2-M^2_k}\,\frac{-i g^{\sigma \sigma '}}{k_1^2-M^2_W}\, \frac{-i g^{\rho \rho '}}{k_2^2-M^2_W}\nonumber\\
 && \xrightarrow{M^2_{k,W} \gg k^2_{1,2},l^2}\quad \overline{v_e}\,\gamma_{\rho }\,\mathrm{P_R}\,\gamma_{\sigma }\,u_\mu\,\frac{i g^2}{2\,M^4_W}\,\underbrace{\sum_{k = 4, 5, ...} \frac{U_{ek} U_{\mu k}}{M^2_k}}_{\equiv \langle M_N^{-1} \rangle_{\mu e}}
\label{eq:MajoranaAmputated}
\end{eqnarray}
from the heavy Majorana neutrino exchange, where the sum extends over all heavy mass eigenstates $N_k$ with admixtures $U_{a k}$ to the active flavours $a$.

At first sight something seems to be wrong, given that the Lorentz structures of Eqs.~\eqref{eq:DoublyAmputated} and~\eqref{eq:MajoranaAmputated} differ. However, taking into account that the hadronic part is symmetric under exchange of the indices, i.e.\ $J^\sigma(\vec{x}_1) J^\rho (\vec{x}_2)=J^\sigma(\vec{x}_2) J^\rho (\vec{x}_1)$, it becomes clear that only the symmetric part of
\begin{equation*}
\gamma_\rho \, \mathrm{P_R} \,\gamma_\sigma = \gamma_\rho \,\gamma_\sigma \, \mathrm{P_L} = \left( g_{\rho \sigma} + \frac{1}{2} \big[\gamma_\rho,\,\gamma_\sigma\big] \right) \mathrm{P_L}
\end{equation*}
contributes to the decay rate. Consequently, the relevant part of Eq.~(\ref{eq:MajoranaAmputated}) is given by
\begin{equation}
 \overline{v}_e\,\mathrm{P_L}u_\mu\,\frac{i g^2\,g_{\sigma \rho}}{2\,M^4_W}\,\langle M_N^{-1} \rangle_{\mu e}\,.
 \label{eq:MajoranaAmputatedSym}
\end{equation}
In contrast to the doubly charged scalar, which only couples to right-hand charged leptons, the heavy Majorana neutrino interacts weakly which leads to left-handed 'external' leptons. Upon calculating the decay rate, both amplitudes are spin-summed and spin-averaged. As a result, only
\begin{equation}
\frac{1}{2}\sum_{\text{spins}} \big | \overline{v}_e\, \mathrm{P_R}\, u_\mu \big|^2 \qquad \text{and} \qquad \frac{1}{2}\sum_{\text{spins}} \big | \overline{v}_e\, \mathrm{P_L}\, u_\mu \big|^2
\end{equation}
are of importance to the final result. Since both expressions equally lead to the factor $1/4$, the chirality of the external leptons does not play a role and can be neglected for the matching. We thus obtain the following correspondence between the two models:
\begin{equation}
\frac{2\,f^*_{e\mu}\,v^4\,\xi}{\Lambda^3\,M^2_S} \qquad \Longleftrightarrow \qquad \langle M_N^{-1} \rangle_{\mu e},
\end{equation}
which can be understood when comparing Eqs.~\eqref{eq:DoublyAmputated} and~\eqref{eq:MajoranaAmputatedSym}.

\subsection{\label{sec:matching_general}Matching particle physics models onto the corresponding operator coefficients}

Now that we have compared two concrete settings to each other, we will show that this is not a mere coincidence and demonstrate on a general ground how to match a model onto the general effective vertex coefficients presented in Eq.~\eqref{eq:short-range}, and thus justify Sec.~\ref{sec:matching_simple}.

\begin{figure}[h]
 \centering
 \begin{minipage}{4.5cm}
  \includegraphics[width=5.5cm]{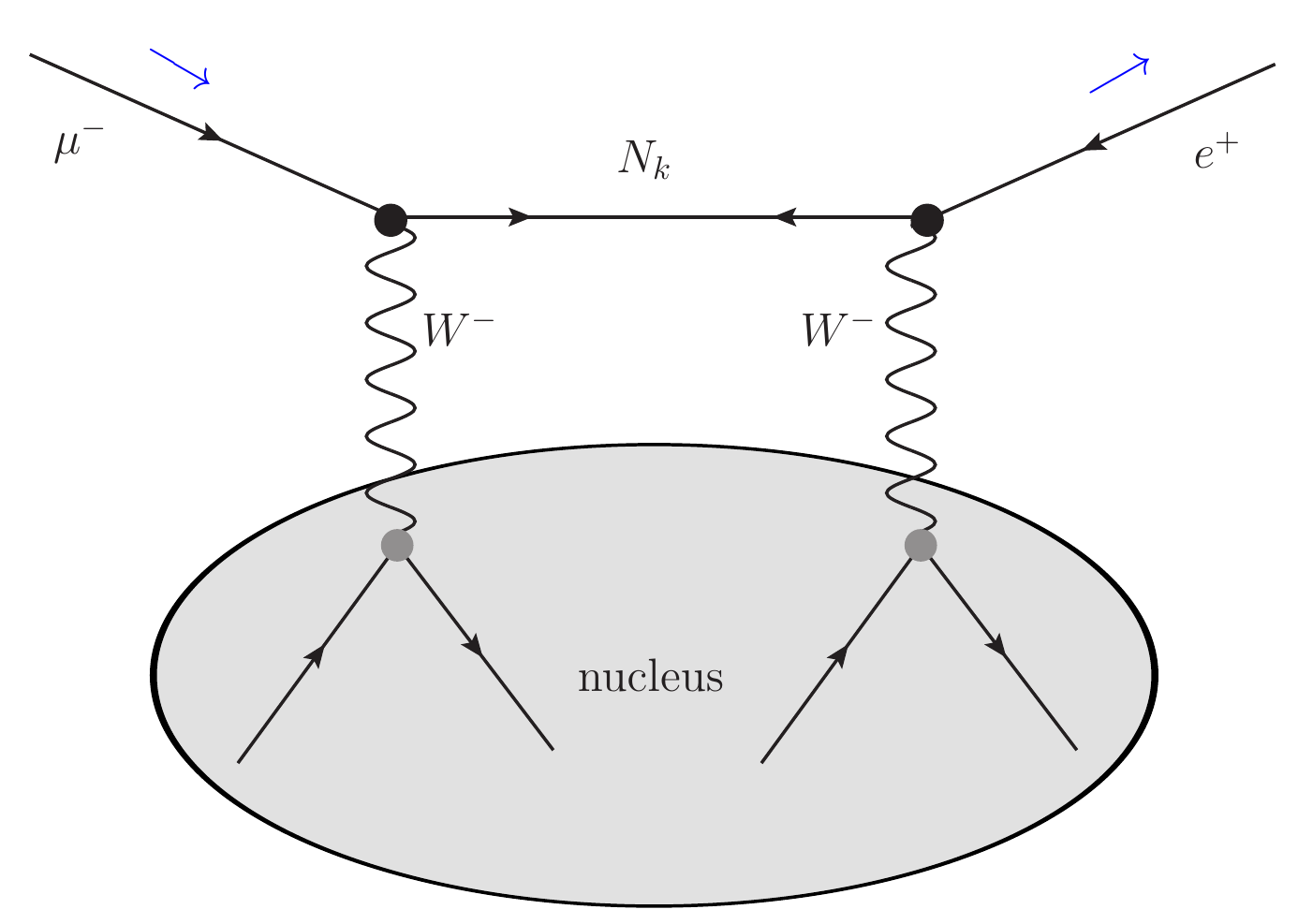}
 \end{minipage}
 \hspace{0.7cm}
 $\Longleftrightarrow$
 \hspace{0.7cm}
 \begin{minipage}{7cm}
  \includegraphics[width=9cm]{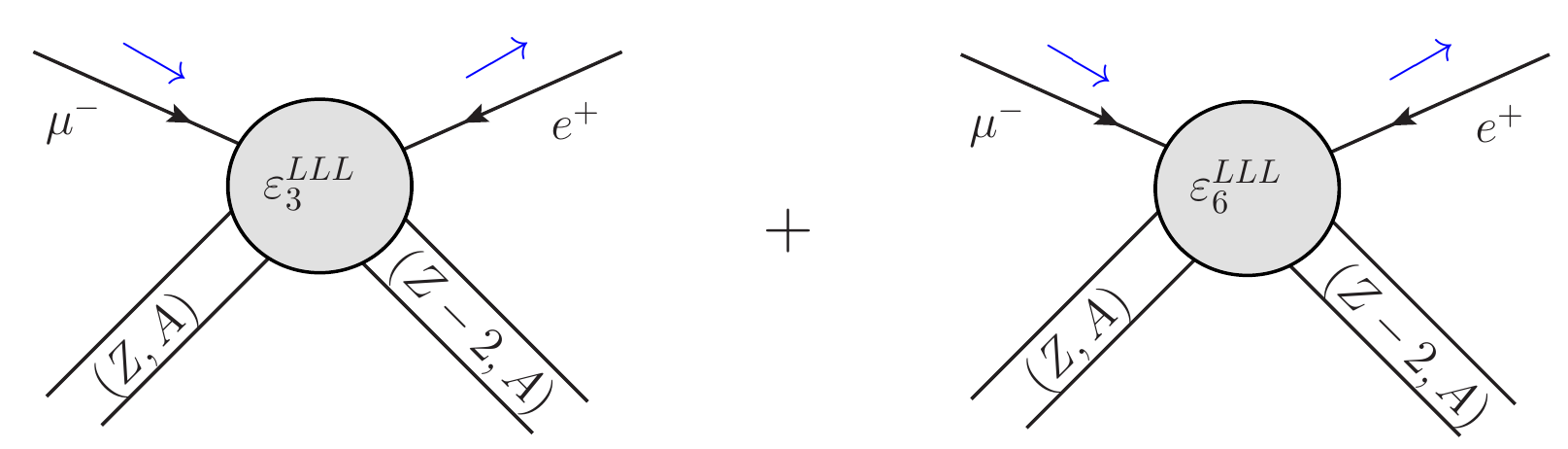}
 \end{minipage}
 \caption{\label{fig:MajoranaMatch}Model with heavy Majoranas $N_k$ mapped onto short-range operators.}
\end{figure}

We start with the model already used in the previous subsection, in which the SM particle content is extended by a number $n$ of SM singlet, right-handed Majorana neutrinos $N_k$, as used in Ref.~\cite{Domin:2004tk}. When rotating the full neutrino mass matrix to a diagonal shape, we end up with $k = 3 + n$ Majorana neutrinos with masses $m_k$. Choosing the specific setting of a seesaw model~\cite{Minkowski:1977sc,Yanagida:1979as,GellMann:1980vs,Glashow:1979nm,Mohapatra:1979ia,Schechter:1980gr}, one obtains three very light (active) neutrinos $\nu_{1,2,3}$ and $n$ heavy (sterile) neutrinos $N_{4, 5, ...}$. Vice versa, the SM's neutrino flavour eigenstates $\nu_a$ can be expressed in terms of light and heavy Majorana mass eigenstates,
\begin{equation*}
 \nu_a = \sum_{l=1, 2, 3\ (\rm light)} U_{al} \nu_l + \sum_{l=4, 5, ...\ (\rm heavy)} U_{al} N_l\,,
\end{equation*}
which allows for the suppressed coupling of charged leptons to heavy Majorana neutrinos, see Fig.~\ref{fig:FeynRule2}, whose strength is parametrised by the active-sterile mixing element $U_{al}$ (where $l = 4, 5, ...$). This coupling results in the realisation of the $\mu^-$-- $e^+$ conversion, as depicted on the left-hand side of Fig.~\ref{fig:MajoranaMatch}. Note that, within this model, the analogous process with light instead of heavy neutrinos leads to a contribution to the long-range part of $\epsilon_3$, which we disregard for the time being.

From the left-hand side of Fig.~\ref{fig:MajoranaMatch}, we obtain
\begin{equation}
\begin{split}
 i\mathcal{M}_k=&\,\overline{d}\,\frac{i g}{2\sqrt{2}}\,V_{ud}\,\gamma^\nu\,\big(1-\gamma_5\big)\,u\,\overline{d}\,\frac{i g}{2\sqrt{2}}\,V_{ud}\,\gamma^\rho\,\big(1-\gamma_5\big)\,u \,\frac{-i g_{\nu \nu'}}{l^2_1-M^2_W}\,\frac{-i g_{\rho \rho'}}{l^2_2-M^2_W}\\
 & \cdot \overline{e^c}\,\frac{-i g}{2\sqrt{2}}\,U_{ek}\,\gamma^{\nu'}\,\big(1+\gamma_5 \big)\,\frac{i \big(\slashed{q}+M_k\big)}{q^2-M^2_k}\,\frac{i g}{2\sqrt{2}}\,U_{\mu k}\,\gamma^{\rho'}\,\big(1-\gamma_5\big)\,\mu\,,
 \end{split}
\end{equation}
with a sum over the different $k$ in case more than one heavy neutrino exist. In the short-range limit (i.e., $l^2_{1,2},\,q^2 \ll M^2_k,\,M^2_W$), and by summing over all heavy mass eigenstates, this turns into
\begin{equation}
 i\mathcal{M}= i \frac{g^4}{64\,M^4_W}\,V^2_{ud}\,\langle M^{-1}_N \rangle_{\mu e}\,\overline{d}\,\gamma^\nu\,\big(1-\gamma_5\big)\,u\,\overline{d}\,\gamma^\rho\,\big(1-\gamma_5\big)\,u\,\overline{e^c}\,\gamma_\nu \, \big(1+\gamma_5\big)\,\gamma_\rho \,\big(1-\gamma_5\big)\,\mu\,.
\end{equation}
The leptonic current is expressed in terms of the bilinear covariants to match Eq.~\eqref{eq:short-range},
\begin{equation}
 \overline{e^c}\,\gamma_\nu \, \big(1+\gamma_5\big)\,\gamma_\rho \,\big(1-\gamma_5\big)\,\mu= 2 g_{\nu \rho}\, \underbrace{\overline{e^c}\,\big(1-\gamma_5\big)\,\mu}_{= j_L}  + 2i\, \underbrace{\overline{e^c}\,\sigma_{\nu \rho}\,\big(1-\gamma_5\big)\,\mu\,}_{= j_{L,\nu \rho}}.
\end{equation}
In terms of Fermi's constant and using hadronic and leptonic currents, this amplitude then takes the form
\begin{equation}
 i\mathcal{M}=i\,\frac{G^2_F}{2m_p}\Big[ \underbrace{2 U^2_{ud}\,m_p\,\langle M^{-1}_N \rangle_{\mu e}}_{=\epsilon^{LLL}_3}J_L^\nu\,J_{L,\nu}\,j_L+  \underbrace{2i V^2_{ud}\,m_p\,\langle M^{-1}_N \rangle_{\mu e}}_{=\epsilon^{LLL}_6}J_L^\nu\,J^\rho_L\,j_{L,\nu \rho} \Big]\,.
\end{equation}
As already indicated on the right-hand side of Fig.~\ref{fig:MajoranaMatch}, the structures $\epsilon^{LLL}_{3,6}$ are realised in this model. Due to the symmetry of the non-relativistic hadronic currents, however, we can simply omit $\epsilon_6$. Thus, in the end, the seesaw model only admits the single operator
\begin{equation}
\frac{G_F^2}{2 m_p}\,\epsilon^{LLL}_3\, J_L^\nu\,J_{L,\nu}\,j_L\,.
\end{equation}

\begin{figure}
 \centering
 \begin{minipage}{4.5cm}
  \includegraphics[width=5.5cm]{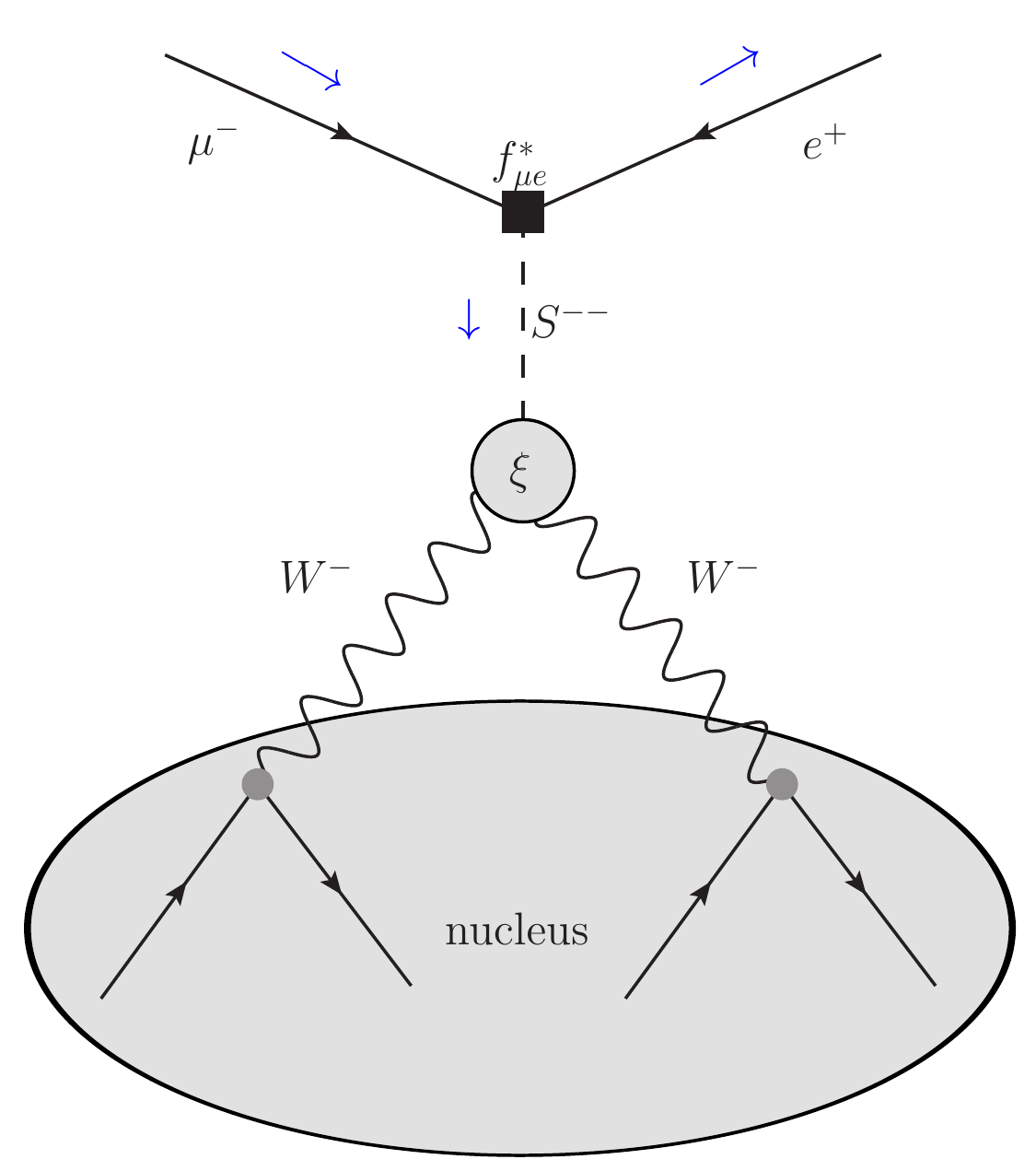}
 \end{minipage}
 \hspace{1cm}
 $\Longleftrightarrow$
 \hspace{1cm}
 \begin{minipage}{7cm}
  \includegraphics[width=6cm]{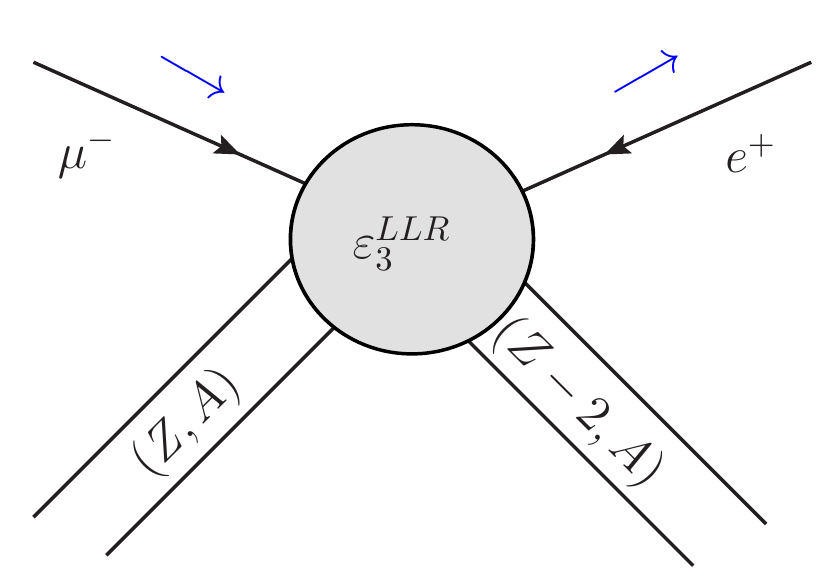}
 \end{minipage}
 \caption{Model with doubly charged scalar $S^{--}$ mapped onto a short-range operator. }
 \label{fig:ScalarMatch}
\end{figure}

Another model that includes LNV is the SM extended by a doubly charged scalar. This model was introduced in Sec.~\ref{sec:rate-explicit_example}. Within this setting, the LNV $\mu^-$-- $e^+$ conversion is realised by the left-hand side of Fig.~\ref{fig:ScalarMatch}. 
\\
The corresponding amplitude is given by
\begin{equation}
\begin{split}
 i\mathcal{M}=&\,\overline{d}\,\frac{i g}{2\sqrt{2}}\,V_{ud}\,\gamma^\nu\,\big(1-\gamma_5\big)\,u\,\overline{d}\,\frac{i g}{2\sqrt{2}}\,V_{ud}\,\gamma^\rho\,\big(1-\gamma_5\big)\,u \,\frac{-i g_{\nu \nu'}}{l^2_1-M^2_W}\,\frac{-i g_{\rho \rho'}}{l^2_2-M^2_W}\\
 &\overline{e^c}\,2i f^*_{e \mu}\,\frac{1}{2}\,\big(1+\gamma_5\big)\,\mu \frac{-i g^2 v^4\,\xi}{2\Lambda^3}\,g^{\rho' \nu'}\,\frac{i}{q^2-M^2_S}\,.
 \end{split}
\end{equation}
Taking the short-range limit ($l^2_{1,2},\,q^2 \ll M^2_S,\,M^2_W$), we obtain
\begin{equation}
 i\mathcal{M}=-i\,\frac{G^2_F}{2m_p}\Big[\underbrace{4 V^2_{ud}\,m_p\,\frac{f^*_{e\mu} v^4\,\xi}{\Lambda^3 M^2_S}}_{=\epsilon^{LLR}_3}J_L^\nu\,J_{L,\nu}\,j_R \Big]\,,
\end{equation}
which is precisely the result used in Eq.~\eqref{eq:example-identification}. Note that we have used the hadronic/leptonic currents from Eqs.~\eqref{eq:hadronic-currents} and~\eqref{eq:leptonic-currents}, as well as some standard identifications such as Fermi's constant.

\begin{figure}
 \hspace{-1cm}
 \begin{minipage}{5.5cm}
  \includegraphics[width=6.5cm]{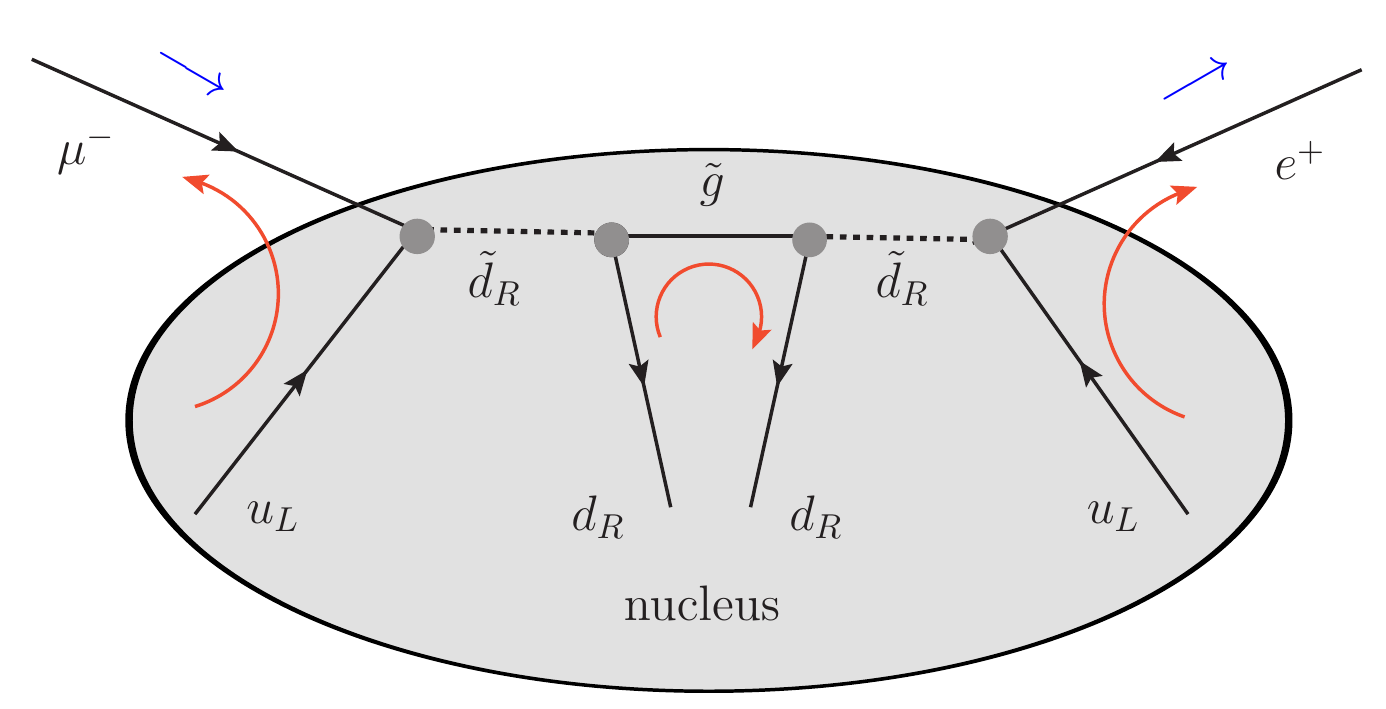}
 \end{minipage}
 \hspace{0.7cm}
 $\Longleftrightarrow$
 \hspace{0.7cm}
 \begin{minipage}{7cm}
  \includegraphics[width=9cm]{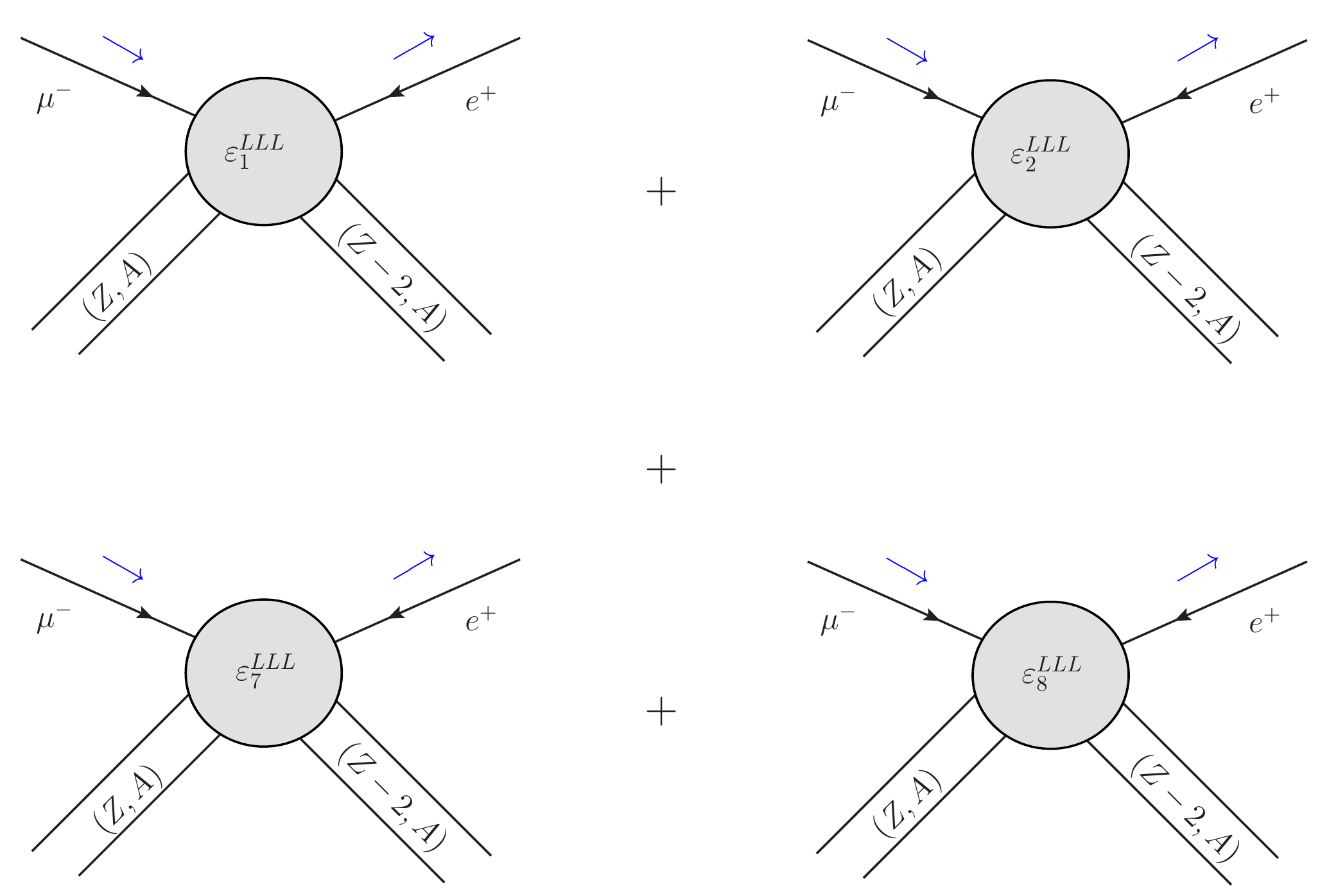}
 \end{minipage}
 \caption{Model with $R$-parity violation: realisation via gluino exchange, mapped onto short-range operators. The red arrows denote the fermion flow, i.e.\ the order in which each fermionic chain is written down.}
 \label{fig:RPVGluino}
\end{figure}

Finally, we want to briefly discuss another class of models that generate LNV, which we have not yet mentioned, to show that it is by far not only the two examples mentioned that are covered by our formalism. The final example are the so-called $R$-parity violating (RPV) supersymmetric (SUSY) theories. Within the framework of RPV-SUSY, there are several mechanisms that provide LNV which are discussed broadly in the literature, see e.g.\ Refs.~\cite{Faessler:1997db,Faessler:2011qw,Bergstrom:2011dt} for the case of $0\nu\beta\beta$. To demonstrate the potential that lies in $\mu^-$-- $e^+$ conversion when contemplating RPV-SUSY, we consider the illustrative case of a gluino exchange being the dominating conversion mechanism. Although there is a number of Feynman diagrams contributing to the $\mu^-$-- $e^+$ conversion due to gluino exchange~\cite{Faessler:2011qw}, we will focus on the diagram given on the left-hand side of Fig.~\ref{fig:RPVGluino} for demonstration purposes. The couplings necessary to realise Fig.~\ref{fig:RPVGluino} can be taken from Ref.~\cite{Barbier:2004ez}, Eq.~(B.8), and~\cite{Faessler:1997db}, Eq.~(18), amongst others. They read:
\begin{equation}
\mathcal{L}= \lambda_{ijk}'\,\tilde{d}^*_{kR}\,\overline{l^c_{iR}}\,u_{jL}+g_3\,\frac{\lambda^{(a)}_{\alpha \beta}}{\sqrt{2}}\,\overline{q^\alpha}\,\mathrm{P_L}\,\tilde{g}^{(a)}\,\tilde{q}_R^\beta\,.
\end{equation}
Here, $\alpha,\,\beta$ denote the colour indices, and $\lambda^{(a)}$ are the Gell-Mann matrices with $a=1,\cdots, 8$.
Upon  employing the orientation of fermion flow as given on the left-hand side of Fig.~\ref{fig:RPVGluino}, we obtain the amplitude
\begin{equation}
i\mathcal{M}\propto \left(\frac{i\,g_3}{\sqrt{2}}\right)^2\,\lambda'_{211}\,\lambda'_{111}\,\frac{i}{p_1^2-m_{\tilde{d}R}^2}\,\frac{i}{p_2^2-m_{\tilde{d}R}^2}\,\overline{(\mu_L)^c}\,u_L\, \overline{(e^c)_R}\,u_L\,\overline{d_R}\,\frac{i(\slashed{p}_1-\slashed{p}_2+M_{\tilde{g}})}{(p_1-p_2)^2-M^2_{\tilde{g}}}\,\big(d_R\big)^c\,.
\end{equation}
Taking the short-range limit, where $p_1^2,\, p_2^2,\,(p_1-p_2)^2 \ll m_{\tilde{d}_R}^2,\,M^2_{\tilde{g}}$, the amplitude takes the form
\begin{equation}
i\mathcal{M} \propto \frac{i g^2_3}{2 m_{\tilde{d}R}^4\,M_{\tilde{g}}}\,\lambda'_{211}\,\lambda'_{111} \Big(\overline{(\mu_L)^c}\,u_L\Big)\Big( \overline{(e^c)_R}\,u_L\Big)\Big(\overline{d_R}\,\big(d_R\big)^c \Big)\,.
\end{equation}
To match this expression onto the operators in Eq.~\eqref{eq:short-range}, we need to rearrange the fermionic fields which can be done by employing Fierz transformations and some algebraic acrobatics. From rearranging the fermionic fields, we obtain that the following four effective operators:
\begin{eqnarray}
&&\Big(\overline{(\mu_L)^c}\,u_L\Big)\Big( \overline{(e^c)_R}\,u_L\Big)\Big(\overline{d_R}\,\big(d_R\big)^c \Big)\nonumber\\
&=& - \frac{1}{2}\,\underbrace{\Big(\overline{d}\,\mathrm{P_L} u\Big)\Big(\overline{d}\,\mathrm{P_L} u\Big)\Big(\overline{e^c}\,\mathrm{P_L} \mu\Big)}_{\to \epsilon^{LLL}_1} - \underbrace{\Big(\overline{d}\,\sigma_{\nu\rho}\,\mathrm{P_L} u\Big)\Big(\overline{d}\,\sigma^{\nu\rho}\,\mathrm{P_L} u\Big)\Big(\overline{e^c}\,\mathrm{P_L} \mu\Big)}_{\to \epsilon^{LLL}_2}\\
&& - \frac{1}{2}\,\underbrace{\Big(\overline{d}\,\sigma_{\nu\rho}\,\mathrm{P_L} u\Big)\Big(\overline{d}\,\mathrm{P_L} u\Big)\Big(\overline{e^c}\,\sigma^{\nu\rho}\,\mathrm{P_L} \mu\Big)}_{\to \epsilon^{LLL}_7} - i\,\underbrace{\Big(\overline{d}\,\sigma_{\nu\rho}\,\mathrm{P_L} u\Big)\Big(\overline{d}\,\sigma^{\nu\kappa}\,\mathrm{P_L} u\Big)\Big(\overline{e^c}\,\sigma^{\rho}_{\,\,\,\kappa}\,\mathrm{P_L} \mu\Big)}_{\to \epsilon^{LLL}_8}\,.\nonumber
\end{eqnarray}
Note that, due to the non-relativistic treatment of the hadronic currents, the operators corresponding to $\epsilon^{LLL}_{7,8}$ will not contribute to the decay rate at this level of approximation. 

As already demonstrated in~\cite{Geib:2016atx}, the strength of this formalism lies in its factorisation of the nuclear physics from the specifics of the particle physics model realising the conversion process. Consequently, by computing only a small number of NMEs, a wide range of particle physics models can be investigated. However, at this moment, only the NME for $\epsilon_3$ is available and we are in need of further NME computations for future analysis of e.g.\ RPV-SUSY models.

\section{\label{sec:conc}Conclusions}

In this paper, we have presented the complete computation of the rate for the lepton flavour \emph{and} number violating $\mu^-$-- $e^+$ conversion, mediated by the effective operator $J_x^\nu J_{y,\nu} j_z$. After introducing the effective operator language in the way appropriate for this process, we have detailed the whole pathway from the amplitude to the decay rate. Our main target group are particle physicists, which is why we had a particular focus on displaying the steps related to the nuclear physics part involved as explicitly as possible. We have furthermore pointed out several concrete New Physics realisations of the effective operator used, all of which can in principle be experimentally probed by $\mu^-$-- $e^+$ conversion.

At the moment, with hardly any NME values being available, this is about as far as one could possibly go when aiming to obtain concrete numbers. However, given that we have now detailed how to perform the computation for the operator $\epsilon_3$, it should at least in principle be clear how to approach the computation for other effective operators. Furthermore, several nuclear physics theory groups have already shown interest in the process, and if they succeed in obtaining further NMEs, the results from both sides could readily be put together, to see which types of New Physics are the most promising in what concerns this $\mu^-$-- $e^+$ conversion.

We want to end this text by stressing that the investigation of the $\mu^-$-- $e^+$ conversion process has got quite some potential to it. It is a rare occasion in physics that we can expect near-future experiments to realistically improve a limit by four to five orders of magnitude. This is an opportunity we should not ignore, which is why we hope to have provided one of the initial sparks for further and more detailed investigations.

\section*{Acknowledgements}

We would like to thank D.~Dercks, F.~Simkovic, and K.~Zuber for useful discussions. AM acknowledges partial support by the the Micron Technology Foundation, Inc., as well as by the European Union's Horizon 2020 research and innovation programme under the Marie Sklodowska-Curie grant agreements No.~690575 (InvisiblesPlus RISE) and No.~674896 (Elusives ITN).

\appendix

\section{\label{app:A}Appendix: Differences between our notation and that from Ref.~\cite{Domin:2004tk}}
\renewcommand{\theequation}{A-\arabic{equation}}
\setcounter{equation}{0}  

To compare our results for the amplitude and subsequently the decay rate for $\mu^+$-- $e$ conversion with those from Ref.~\cite{Domin:2004tk}, further remarks are in order:
\begin{enumerate}

\item First, note that the normalisation of free spinors in~\cite{Domin:2004tk} differs from the one used in the derivation above. The normalisation of spinors as used in Ref.~\cite{Domin:2004tk} is stated in their Appendix~A, and it corresponds to the following spin sums:
\begin{equation}
 \sum_r \big|u^r (p) \overline{u^r}(p) \big|^2= \slashed{p}+m \mathds{1} \qquad \text{and} \qquad \sum_r \big| v^r (p) \overline{v^r}(p)\big|^2=\slashed{p}-m\mathds{1}\,.
\end{equation}
We, on the other hand, use a normalisation that leads to
\begin{equation}
 \sum_r \big|u^r (p) \overline{u^r}(p) \big|^2= \frac{1}{4E}\big(\slashed{p}+m \mathds{1}\big) \qquad \text{and} \qquad \sum_r \big| v^r (p) \overline{v^r}(p)\big|^2=\frac{1}{4E}\big(\slashed{p}-m \mathds{1}\big)\,.
\end{equation}
Although the resulting decay rate will not differ, we must translate $u_\mu \to 1/\sqrt{2E_\mu}\, u_\mu$ and $\overline{v}_e \to 1/\sqrt{2E_e}\, \overline{v}_e$ in order to compare on the matrix-element level. Note that the spinor expression in Eq.~(A3) of~\cite{Domin:2004tk} uses the non-relativistic limit in the Dirac representation.

\item Second, it is important to take into account that one will encounter an additional factor of of $1/(2\pi)^{3/2}$ in the matrix element of Ref.~\cite{Domin:2004tk}, which arises from preponing the phase space integral factor of $1/(2\pi)^3$ to the matrix element, which we do not.

\item Third, Ref.~\cite{Domin:2004tk} only introduces the relativistic Coulomb factor $F(Z-2,E_e)$ that accounts for the positron propagating under the influence of the nuclear field when stating the decay rate. We, however, already introduce it when deriving the amplitude.

\item Fourth, we have to recall that Ref.~\cite{Domin:2004tk} realises $\mu^-$-- $e^+$ conversion by means of a heavy Majorana, whereas we depend on the doubly charged scalar. Thus, we need to perform some sort of matching in order to replace the model-dependent doubly charged scalar contribution by the respective heavy Majorana part, as discussed in detail in Sec.~\ref{sec:matching_simple}. From there, we obtain the relation
\begin{equation}
\frac{2\,f^*_{e\mu}\, V_{ud}^2\, v^4\,\xi}{\Lambda^3\,M^2_S} \qquad \Longleftrightarrow \qquad \frac{\epsilon_3^{LLx}}{2 m_p} \qquad \Longleftrightarrow \qquad \langle M_N^{-1} \rangle_{\mu e}\,,
\end{equation}
with $x=L,R$ and $\langle M_N^{-1} \rangle_{\mu e}=\sum_{k = 4, 5, ...} U_{ek} U_{\mu k}/M^2_k$, where $U_{a k}$ denotes the admixture of the heavy neutrino mass eigenstate $N_k$ to the active flavour $a$.

\item Last but not least, the chiralities of the external charged leptons have to be considered. While in our scenario both muon and positron are right-handed, the charged leptons are left-handed when coupling to the heavy Majorana neutrinos. Furthermore, note that Ref.~\cite{Domin:2004tk} uses another convention for the $\gamma$-matrices which is based on employing the Pauli metric~\cite{Bilenky:1987ty}\footnote{It is useful to check footnote~3 on page~676.} instead of the Minkowski metric. This results into the left-/right-handed projectors having a reversed sign in front of $\gamma_5$ with reference to their definition in the basis we use, namely $g^{\nu\rho}=\{+,-,-,-\}$. Since the spin sums lead to the same factor independent of the charged leptons being left- or right-handed, we can consistently replace
\begin{equation}
 \overline{v}_e(k_e)\,\mathrm{P_R}\,u_\mu (k_\mu) \xrightarrow{S^{++}\to N_k}\overline{v}_e(k_e)\,\mathrm{P_L}\,u_\mu (k_\mu) \xrightarrow{\text{adapt notation}} \frac{1}{\sqrt{4 E_e E_\mu}}\,\frac{1}{2}\,\overline{v}_e(k_e)\,\big(1+\gamma_5\big)\,u_\mu (k_\mu)
\end{equation}
for the sake of comparing our computation to the amplitude in~\cite{Domin:2004tk}.\footnote{Here, in the very last step, we switched to \cite{Domin:2004tk}'s normalisation of free spinors and to the Pauli metric notation.}

\end{enumerate}
Combining the above comments, our matrix element from Eq.~\eqref{eq:Amplitude1} translates into
\begin{equation}
\begin{split}
\mathcal{M}=&\,\bigg(\frac{G_F}{\sqrt{2}}\bigg)^2\,\langle M_N^{-1} \rangle_{\mu e}\,\frac{g_A^2 m_p m_e}{R}\,\delta(E_f-E_i+E_e-E_\mu)\,\mathcal{M}^{(\mu^-,e^+)\,\phi}\\
&\frac{1}{(2\pi)^{3/2}}\,\frac{1}{\sqrt{4 E_e E_\mu}}\,\overline{v}_e(k_e)\,\big(1+\gamma_5\big)\,u_\mu (k_\mu)\,,
\end{split}
\end{equation}
which agrees with Eq.~(32) of~\cite{Domin:2004tk}.

\section{\label{app:B}Appendix: Understanding Eq.~(49) from Ref.~\cite{Domin:2004tk}}
\renewcommand{\theequation}{B-\arabic{equation}}
\setcounter{equation}{0}  

To understand the notation in Eq.~(49) of Ref.~\cite{Domin:2004tk}, we use this appendix to demonstrate in detail how to rewrite the exponential functions from Eq.~\eqref{eq:NME1}:
\begin{equation}
 \int \diff k \,k^2 \diff \Omega_k\,\langle N' \big |\,\mathrm{e}^{i \vec{k}\cdot (\vec{r}_m-\vec{r}_l)}\mathrm{e}^{-i \vec{k}_e \cdot \vec{r}_l}\,\big| N \rangle\,,
 \label{eq:Angular1}
\end{equation}
with $k=| \vec{k}|$. Since we only take care of the angular integration in the following, we dropped some $\vec{k} ^2$-dependent parts of Eq.~(\ref{eq:NME1}). Introducing new coordinates,
\begin{equation}
 \vec{r}_{lm}=\vec{r}_l-\vec{r}_m \quad \text{with} \quad r_{lm}=\big| \vec{r}_{lm} \big|\, , \quad \text{and} \quad\vec{R}_{lm}=\frac{\vec{r}_l+\vec{r}_m}{2} \quad \text{with} \quad R_{lm}=\big| \vec{R}_{lm} \big|\,,
\end{equation}
we perform the angular integration,
\begin{equation}
 \int \diff \Omega_k\,\mathrm{e}^{i \vec{k}\cdot (\vec{r}_m-\vec{r}_l)}= 4\pi j_0 \big(k r_{lm}\big)\,,
\end{equation}
and replace the remaining exponential function by its plane wave decomposition:
\begin{equation}
 \mathrm{e}^{-i\vec{k}\cdot \vec{x}}=\big(\mathrm{e}^{i\vec{k}\cdot \vec{x}}\big)^*=4\pi \sum_{\lambda=0}^{\infty} (-i)^\lambda j_\lambda \big(k x\big) \sum_{m_\lambda=-\lambda}^{\lambda} Y_\lambda^{m_\lambda} \big(\vartheta_k, \varphi_k \big)\,Y_\lambda^{m_\lambda\,*} \big(\vartheta_x, \varphi_x \big)\,,
\end{equation}
where $k=| \vec{k}\,|$, and $x=| \vec{x}\,|$. Here, $\vartheta_{k,x}$ and $\varphi_{k,x}$ are the azimuthal, and polar angles which fix the directions of $\vec{k}$ and $\vec{x}$, respectively. Furthermore, $j_\lambda$ denotes the spherical Bessel function and $Y_\lambda^{m_\lambda}$ the spherical harmonic. Thus, Eq.~\eqref{eq:Angular1} takes the form
\begin{equation}
\begin{split}
 (4\pi)^3 &\int \diff k \,k^2 \,\langle N' \big |\,j_0 \big(k r_{lm}\big)\,\sum_{\lambda, \lambda'} (-i)^{\lambda+\lambda'} j_{\lambda}\big(k_e R_{lm}\big)j_{\lambda'}\big(k_e r_{lm}/2\big)\\
 &\sum_{m_\lambda,m_{\lambda'}}\,Y_\lambda^{m_\lambda} \big(\vartheta_{k_e}, \varphi_{k_e} \big)\,{Y_\lambda^{m_\lambda}}^* \big(\vartheta_{R_{lm}}, \varphi_{R_{lm}} \big)\,Y_{\lambda'}^{m_{\lambda'}} \big(\vartheta_{k_e}, \varphi_{k_e} \big)\, {Y_{\lambda'}^{m_{\lambda'}}}^* \big(\vartheta_{r_{lm}}, \varphi_{r_{lm}} \big)\,\big| N \rangle\,.
 \end{split}
\end{equation}
Next, we employ the well-known addition theorem for Legendre polynomials,
\begin{equation}
P_\lambda \big( \cos \vartheta_{n n'} \big)=\frac{4\pi}{2\lambda +1}\,\sum_{m_\lambda=-\lambda}^{\lambda} Y_\lambda^{m_\lambda} \big(\vartheta_n, \varphi_n \big)\,Y_\lambda^{m_\lambda\,*} \big(\vartheta_{n'}, \varphi_{n'} \big)\,,
\end{equation}
where $\cos \vartheta_{n n'} = \vec{n}\cdot \vec{n}' = \cos \vartheta_{n} \cos \vartheta_{n'} + \sin \vartheta_{n} \sin \vartheta_{n'} \cos \big(\varphi_{n}-\varphi_{n'}\big)$. Note that $\vec{n}$ and $\vec{n}'$ are unit vectors. The resulting Legendre polynomials $P_\lambda$ can themselves be phrased in terms of spherical harmonics
\begin{equation}
 P_\lambda \big( \cos \vartheta_{n n'} \big)=\sqrt{\frac{4\pi}{2\lambda+1}}\,Y_\lambda^0 \big(\vartheta_{n n'}, \varphi_{n n'} \big)\,,
\end{equation}
and we hence obtain:
\begin{equation}
 \begin{split}
 (4\pi)^2 &\int \diff k \,k^2 \,\langle N' \big |\,j_0 \big(k r_{lm}\big)\,\sum_{\lambda, \lambda'} (-i)^{\lambda+\lambda'}\,\sqrt{(2\lambda +1)(2\lambda'+1)}\, j_{\lambda}\big(k_e R_{lm}\big)j_{\lambda'}\big(k_e r_{lm}/2\big)\\
 &Y_\lambda^{0} \big(\Omega_{k_e R_{lm}} \big)\,Y_{\lambda'}^{0} \big(\Omega_{k_e r_{lm}} \big)\,\big| N \rangle\,,
 \end{split}
\end{equation}
with $\Omega_{k_e R_{lm}}\equiv \big(\vartheta_{k_e R_{lm}}, \varphi_{k_e R_{lm}} \big)$ and $\Omega_{k_e r_{lm}}\equiv \big(\vartheta_{k_e r_{lm}}, \varphi_{k_e r_{lm}} \big)$.

We can further rephrase the spherical harmonics by using the inverse Clebsch-Gordan relation, see Eq.~(4-b) in complement $C_X$ of~\cite{Cohen:1977}:
\begin{equation}
 Y_\lambda^{m} \big(\Omega_{1} \big)\,Y_{\lambda'}^{m'} \big(\Omega_{2} \big)= \sum_{LM} \big( \lambda m_1 \lambda' m_2 \big| LM \big)\,\underbrace{\Phi_{LM} \big(\Omega_1,\Omega_2\big)}_{\equiv \Big\{ Y_{\lambda} \big(\Omega_{1} \big)\otimes Y_{\lambda'} \big(\Omega_{2}\big)\Big\}_{LM}} \,,
\end{equation}
where the connection to the irreducible tensors is established with help of Eq.~(1) in Chapter~5.16 of Ref.~\cite{Varshalovich:1988}.

One can connect the Clebsch-Gordan coefficients to the $\mathit{3j}$ symbols by means of 
\begin{equation}
 \big( \lambda_1 m_1 \lambda_2 m_2 \big| LM \big)= (-1)^{\lambda_2-\lambda_1-M}\,\sqrt{2L+1}\,\begin{pmatrix} \lambda_1 & \lambda_2 & L \\
 m_1 & m_2 & M \end{pmatrix}\,,
\end{equation}
as stated in Eq.~(1.44) of Ref.~\cite{Suhonen:2007}.

In the case of a coherent transition, the operator has to be a scalar, which enforces $L=0$ (see also~\cite{Varshalovich:1988}, Chapter~3.2.1). Furthermore, the $\mathit{3j}$ symbols satisfy the following properties:
\begin{equation}
 \begin{split}
  m_1+m_2=&M\,,\\
  \begin{pmatrix} \lambda_1 & \lambda_2 & 0 \\
 m_1 & m_2 & 0 \end{pmatrix}=&(-1)^{\lambda_1-m_1}\,\frac{1}{\sqrt{2\lambda_1+1}}\,\delta_{\lambda_1 \lambda_2}\,\delta_{m_1,-m_2}\,,
 \end{split}
\end{equation}
which can be found in~\cite{Suhonen:2007} under Eqs.~(1.41) and~(1.42). Since we have $m_\lambda=m_{\lambda'}=0$, the quantum numbers of the coupled system are fixed to $L=M=0$. That way, the inital expression in Eq.~\eqref{eq:Angular1} can be rearranged to
\begin{equation}
\begin{split}
 (4\pi)^2 &\int \diff k \,k^2 \,\langle N' \big |\,j_0 \big(k r_{lm}\big)\,\sum_{\lambda}\,\sqrt{2\lambda +1}\, j_{\lambda}\big(k_e R_{lm}\big)j_{\lambda}\big(k_e r_{lm}/2\big)\\
 &\Big\{ Y_{\lambda} \big(\Omega_{k_e r_{lm}} \big)\otimes Y_{\lambda} \big(\Omega_{k_e R_{lm}}\big)\Big\}_{00}\,\big| N \rangle,
 \end{split}
 \label{eq:Angular2}
\end{equation}
valid for the case of a coherent nuclear transition.

This can be simplified further when taking into account that according to Ref.~\cite{Manakov:1977}'s Eq.~(4),
\begin{equation}
 P_\lambda \big( \vec{n}\cdot \vec{n}' \big)= (-1)^\lambda\,4\pi\,\Big\{ Y_\lambda \big(\vec{n} \big)\otimes Y_\lambda \big(\vec{n'} \big)\Big\}_{00}=(-1)^\lambda\,4\pi\,\Big\{ Y_\lambda \big(\vartheta_{n}, \varphi_{n} \big)\otimes Y_\lambda \big(\vartheta_{n'}, \varphi_{n'} \big)\Big\}_{00}\,.
\end{equation}
Applying this relation to $\{\cdots\}_{00}$ in Eq.~\eqref{eq:Angular2}, it becomes obvious that this expression does indeed not depend on $\vec{k}_e$ anymore. We can, consequently, discard the dependence on the positron's momentum and state the final form of the angular part for coherent transitions:
\begin{eqnarray}
 &&\int \diff k \,k^2 \diff \Omega_k\,\langle N' \big |\,\mathrm{e}^{i \vec{k}\cdot (\vec{r}_m-\vec{r}_l)}\mathrm{e}^{-i \vec{k}_e \cdot \vec{r}_l}\,\big| N \rangle \xrightarrow{\text{coherent}} \label{eq:AngularFinal}\\ 
 &&(4\pi)^2 \int \diff k \,k^2 \,\langle N' \big |\,j_0 \big(k r_{lm}\big)\,\sum_{\lambda}\,\sqrt{2\lambda +1}\, j_{\lambda}\big(k_e R_{lm}\big)j_{\lambda}\big(k_e r_{lm}/2\big)\,\Big\{ Y_{\lambda} \big(\Omega_{r_{lm}} \big)\otimes Y_{\lambda} \big(\Omega_{R_{lm}}\big)\Big\}_{00}\,\big| N \rangle\,,\nonumber
\end{eqnarray}
where $\Omega_{R_{lm}}$ and $\Omega_{r_{lm}}$ fix the directions of $\vec{R}_{lm}$ and $\vec{r}_{lm}$ independently of $\vec{k}_e$.

\section{\label{app:C}Appendix: Relevant Feynman rules}
\renewcommand{\theequation}{C-\arabic{equation}}
\setcounter{equation}{0}  

\begin{figure}[h]
\begin{minipage}[c]{14cm}
\begin{subfigure}[c]{7cm}
  \includegraphics[width=7cm]{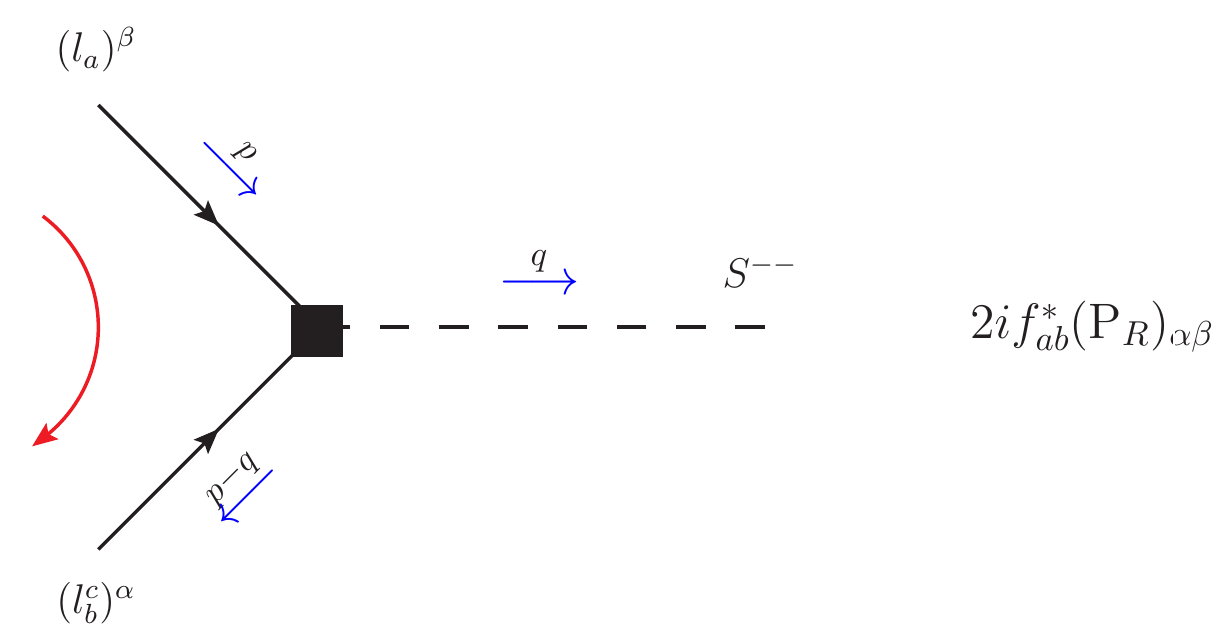}
  \caption{Lepton-lepton-$S^{--}$ interaction with $f^{(*)}_{ab}=f^{(*)}_{ba}$}
  \label{fig:FeynRule1a}
\end{subfigure}
\hspace{2cm}
\begin{subfigure}[c]{7cm}
 \includegraphics[width=5cm]{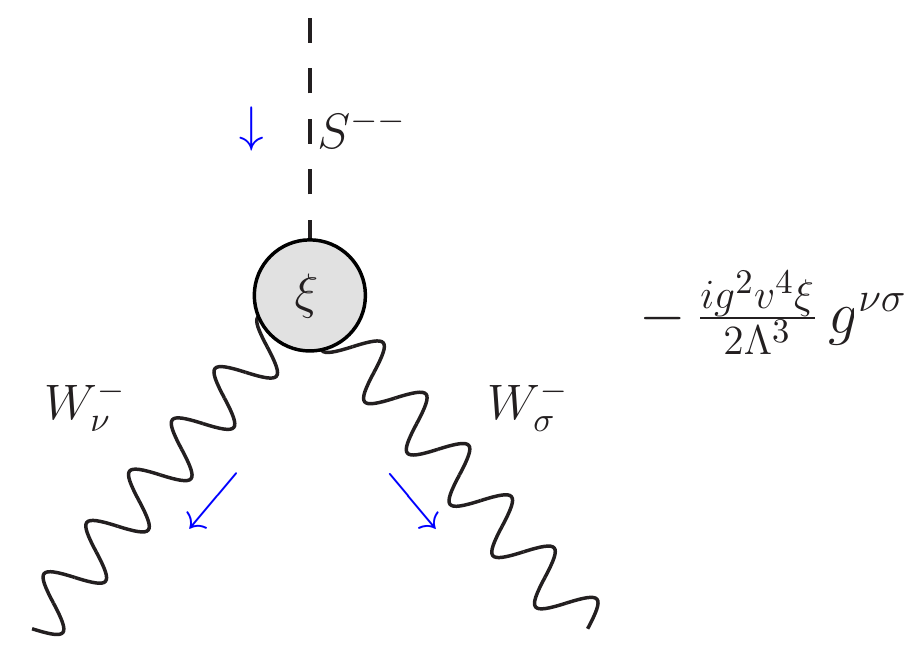}
  \caption{Effective vertex coupling $S^{--}$ to two gauge bosons}
 \label{fig:FeynRule1b}
 \end{subfigure}
\end{minipage}
 \caption{Coupling the doubly charged scalar to the SM}
    \label{fig:FeynRule1}
\end{figure}

\begin{figure}[h]
\begin{minipage}[c]{14cm}
\begin{subfigure}[c]{7cm}
  \includegraphics[width=7cm]{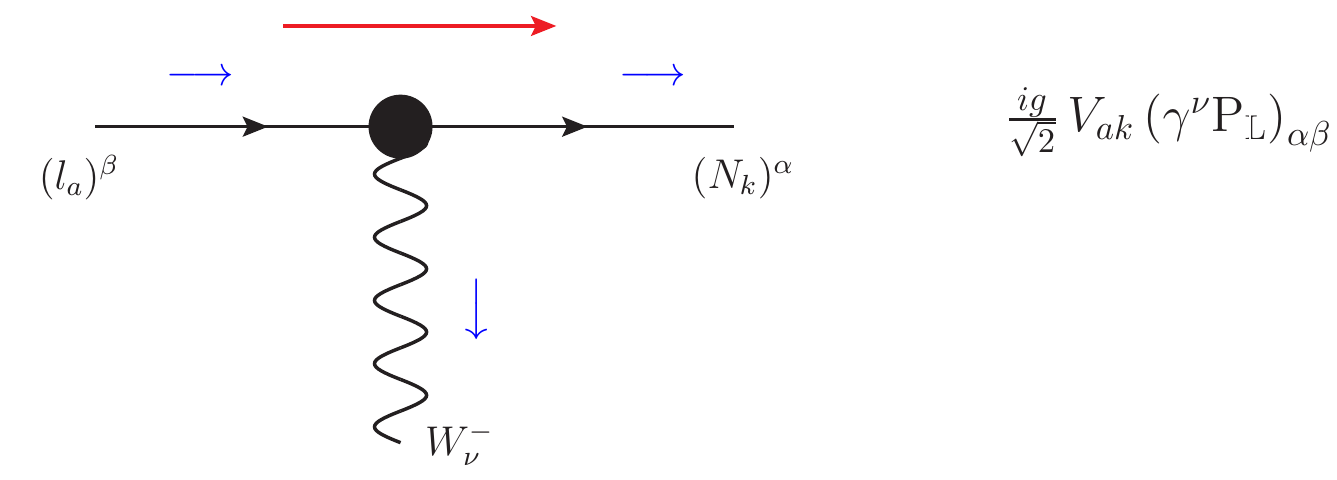}
  \caption{Usual orientation of 'fermion flow'}
  \label{fig:FeynRule2a}
\end{subfigure}
\hspace{2cm}
\begin{subfigure}[c]{7cm}
 \includegraphics[width=7cm]{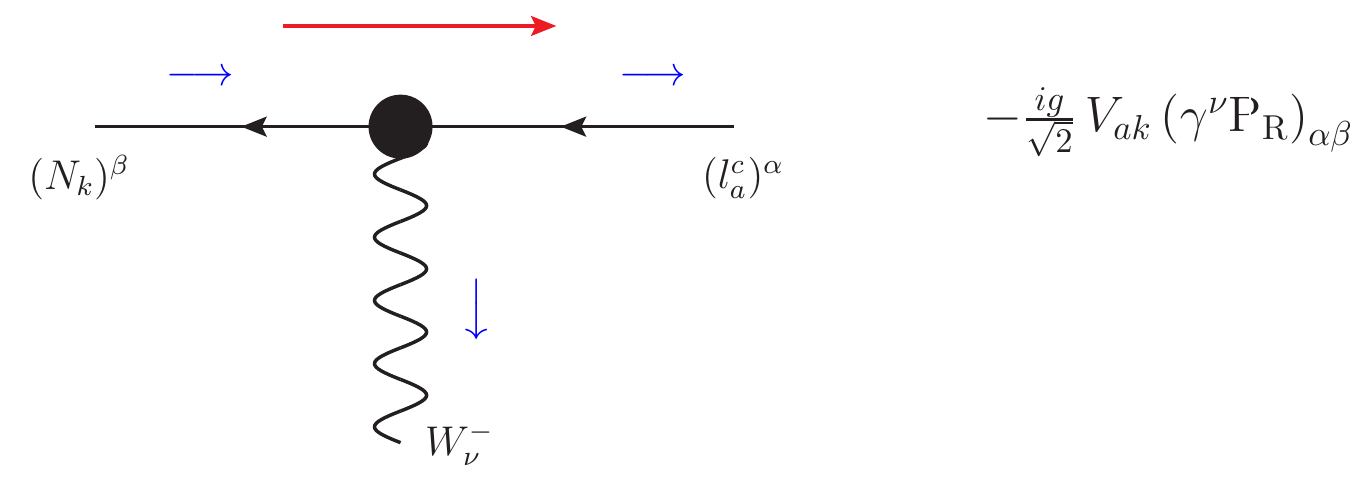}
 \caption{Reversed 'fermion flow'}
   \label{fig:FeynRule2b}
 \end{subfigure}
\end{minipage}
\caption{Coupling the Majorana neutrino to the SM}
\label{fig:FeynRule2}
\end{figure}

\begin{figure}
\centering
 \includegraphics[width=7cm]{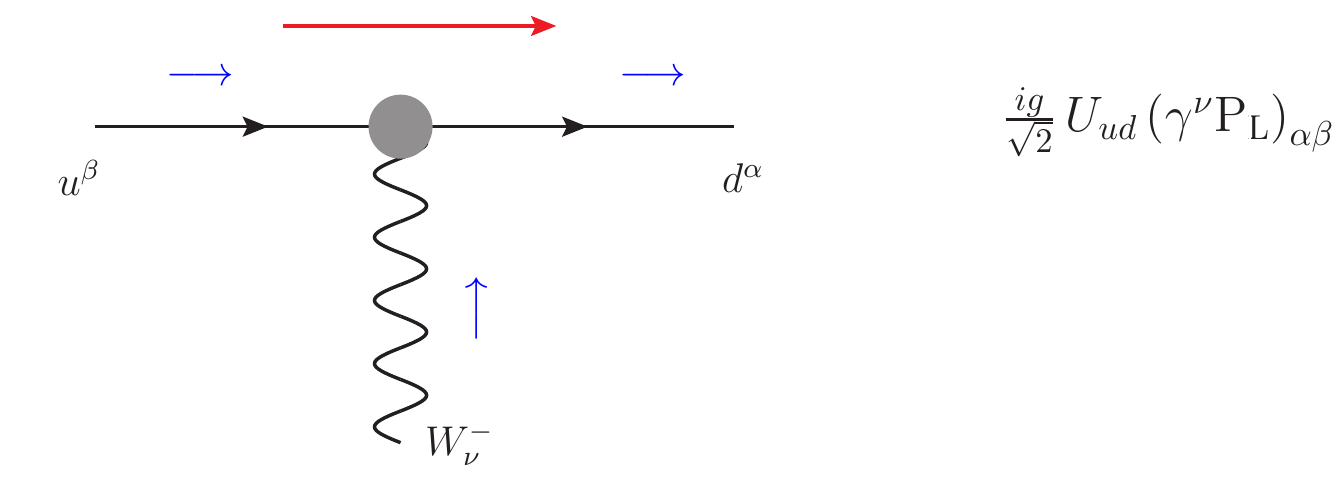}
 \caption{Coupling the $W$-boson to two quarks}
 \label{fig:FeynRule4a}
\end{figure}

\begin{figure}[h]
\begin{minipage}[c]{14cm}
\begin{subfigure}[c]{7cm}
  \includegraphics[width=7cm]{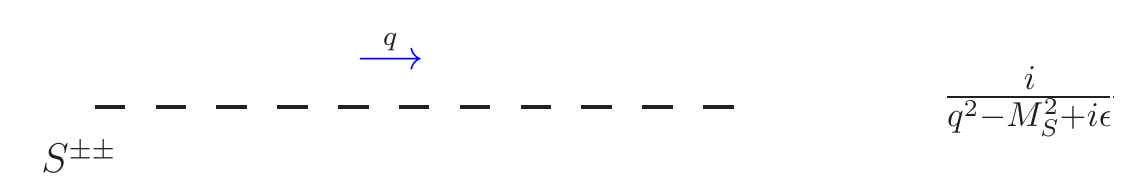}
  \caption{Doubly charged scalar}
  \label{fig:FeynRule3a}
\end{subfigure}
\hspace{2cm}
\begin{subfigure}[c]{7cm}
 \includegraphics[width=7cm]{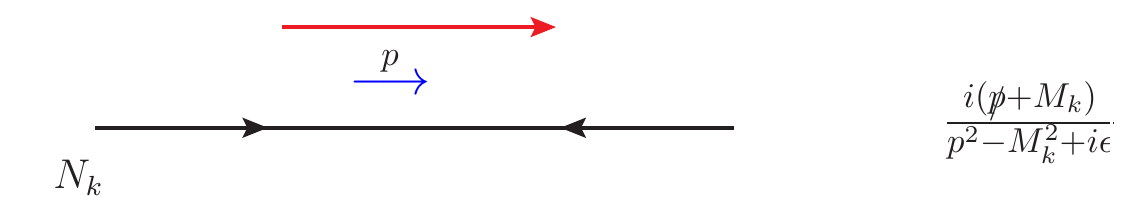}
 \caption{Majorana neutrino}
 \label{fig:FeynRule3b}
\end{subfigure}
\begin{subfigure}[c]{7cm}
 \includegraphics[width=7cm]{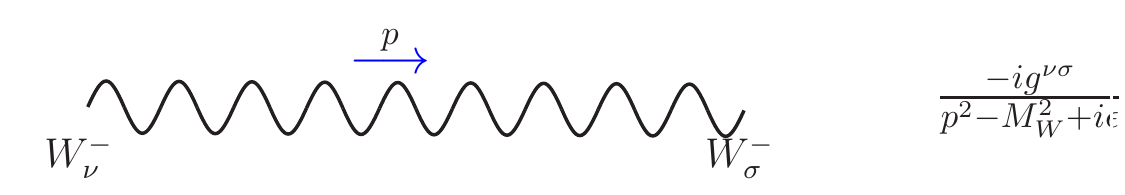}
 \caption{$W$-boson}
 \label{fig:FeynRule3c}
 \end{subfigure}
\end{minipage}
\caption{Propagators}
\label{fig:FeynRule3}
\end{figure}

In order to perform the matching of the model-dependent coefficients in Sec.~\ref{sec:matching}, we make use of the Feynman rules given in Figs.~\ref{fig:FeynRule1} to~\ref{fig:FeynRule3}. Here, $\mathrm{P_{L,R}}$ are the left-/right-handed projectors, the indices $\alpha,\,\beta$ are Dirac spinor indices, the indices $a,\,b=e,\,\mu\,,\tau$ denote the lepton flavour, and $k= 4, 5, ...$ refer to the mass eigenstates of the heavy Majorana neutrinos. We use a model where the conversion is mediated by a doubly charged scalar $S^{\pm \pm}$ which couples to the right-handed charged leptons via a LNV vertex~\cite{King:2014uha}. We aim at comparing this to a model where the conversion is mediated by a heavy Majorana neutrino $N_k$. Note that, because of the fact that there are LNV vertices in our theory, we naturally encounter vertices or Majorana propagators with clashing arrows. For a consistent treatment using the Feynman rule language, we choose a fixed orientation of the ``fermion flow'' for each diagram, i.e.\ the order in which each fermionic chain is written down, and adjust the Feynman rules~\cite{Denner:1992vza,Haber:1984rc,Jones:1983eh}. For example, when reversing the ``fermion flow'' from Figs.~\ref{fig:FeynRule2a} to~\ref{fig:FeynRule2b}, we instead work with the antifield $l^c_a=C\,\overline{l_a}^T$ and alter the Feynman rules accordingly. In Figs.~\ref{fig:FeynRule1} to~\ref{fig:FeynRule3}, the red arrow indicates the orientation of the ``fermion flow'', i.e., of lepton number.

\clearpage
\bibliographystyle{./apsrev}
\bibliography{muon-positron}

\end{document}